\documentclass{aa}
\usepackage{graphicx}
\usepackage{txfonts}
\usepackage[utf8]{inputenc}
\usepackage{color}
\usepackage{multirow}
\graphicspath{{./}{figures/}}

\begin{document}

   \title{Upgraded antennas for pulsar observations in the\\
   Argentine Institute of Radio astronomy}

   \author{G. Gancio\inst{1} \and
          C.~O. Lousto\inst{2} \and L.~Combi\inst{1,2} \and
          S.~del Palacio \inst{1} \and
          F.~G. L\'opez Armengol\inst{1,2} \and
          J.~A. Combi\inst{1,3} \and
          F.~Garc\'ia\inst{4,1} \and
          P.~Kornecki\inst{1} \and
          A.~L. M\"uller\inst{1} \and
          E.~Guti\'errez\inst{1} \and
          F.~Hauscarriaga \inst{1}\and
          G. C.~Mancuso \inst{1}
          }

        \institute{Instituto Argentino de Radioastronom\'{\i}a (IAR), C.C. No. 5, 1894, Buenos Aires, Argentina
        \email{ggancio@iar.unlp.edu.ar}
         \and Center for Computational Relativity and Gravitation, School of Mathematical Sciences, Rochester Institute of Technology, 85 Lomb Memorial Drive, Rochester,New York 14623, US
            \email{colsma@rit.edu}
        \and
             Facultad de Ciencias Astron\'omicas y Geof\'{\i}sicas, Universidad Nacional de La Plata, Paseo del Bosque, B1900FWA La Plata, Argentina
        \and
             Kapteyn Astronomical Institute, University of Groningen, P.O. BOX 800, 9700 AV Groningen, The Netherlands      }

 
 \abstract
   {The Argentine Institute of Radio astronomy (IAR) is equipped with two single-dish 30~m radio antennas capable of performing daily observations of pulsars and radio transients in the southern hemisphere at 1.4 GHz.}
   {We aim to introduce to the international community the upgrades performed and to show that IAR observatory has become suitable for investigations in numerous areas of pulsar radio astronomy, such as pulsar timing arrays, targeted searches of continuous gravitational waves sources, monitoring of magnetars and glitching pulsars, and studies of short time scale interstellar scintillation.}
   {We refurbished the two antennas at IAR to achieve high-quality timing observations. We gathered more than $1\,000$ hours of observations with both antennas to study the timing precision and sensitivity they can achieve.} 
   {We introduce the new developments for both radio telescopes at IAR. We present daily observations of the millisecond pulsar J0437$-$4715 with timing precision better than 1~$\mu$s. We also present a follow-up of the reactivation of the magnetar XTE J1810--197 and the measurement and monitoring of the latest (Feb. 1st. 2019) glitch of the Vela pulsar (J0835--4510).}
   {We show that IAR is capable of performing pulsar monitoring in the 1.4 GHz radio band for long periods of time with a daily cadence. This opens the possibility of pursuing several goals in pulsar science, including coordinated multi-wavelength observations with other observatories. 
   In particular, daily observations of the millisecond pulsar J0437$-$4715 will increase the sensitivity of pulsar timing arrays. We also show IAR's great potential for studying targets of opportunity and transient phenomena such as magnetars, glitches, and fast-radio-burst sources.}

\maketitle
%

\section{Introduction}

   The Argentine Institute of Radio astronomy (IAR; Instituto Argentino de Radioastronom{\'i}a\footnote{\url{http://www.iar.unlp.edu.ar}}) was founded in 1962 as a pioneer radio observatory in South America with two 30-meter parabolic single-dish radio antennas (Fig.~\ref{fig:AntennaII&I}). Antenna 1 (A1) saw its first light in 1966 whereas Antenna 2 (A2) was built later in 1977\footnote{In 2019, A1 and A2 were baptized `Varsavsky' and `Bajaja', respectively, in honor to their contributions to the IAR.}. The IAR's initial purpose was to perform a high sensitivity survey of neutral hydrogen ($\lambda=21$~cm) in the southern hemisphere; this survey ended satisfactorily in the year 2000 with high-impact publications in collaboration with German and Dutch institutions \citep{Testori:2001vp,bajajaetal,Kalberla:2005ts}.
   
   Although the IAR has been a center of intense scientific and technological activity since its foundation, the radio antennas have not been employed in any scientific project since 2001. For the first time in over fifteen years, the IAR antennas are being upgraded to conduct high-quality radio astronomy. The PuMA\footnote{\url{http://puma.iar.unlp.edu.ar}} (Pulsar Monitoring in Argentina) team is a collaboration of scientists and technicians from the IAR and the Rochester Institute of Technology (RIT). The collaboration has been working for 2 years with both antennas, including the implementation of a dedicated backend, the construction of a brand new frontend for A2, and formation of human resources for observations, data analysis, and pulsar astrophysics. This project represents the first systematic pulsar timing observations in South America and the beginning of pulsar science in Argentina. 
   
   In this work, we present the new hardware developments and observations of IAR's radio antennas from the last two years. In Section~\ref{sec:IAR} we give an overview of the current state of the radio observatory, its radio interference environment (RFI), the atomic clock availability, and the new developments for the acquisition software. In Section~\ref{sec:obs} we describe the observational techniques and capabilities, both hardware and observational cadence, as well as their automation. Calibration of total flux densities and polarization are part of our current developments.

   In Section~\ref{sec:science} we give a description of the various scientific projects that are being carried out or will be in the near future with further hardware improvements. One of the major goals of the PuMA collaboration involves high-cadence monitoring of millisecond pulsars (MSPs). In particular, monitoring of PSR J0437$-$4715 (one of the closest MSPs) is of great importance for gravitational wave detection using pulsar timing array techniques.  Other close-by MSPs are also a target of the LIGO-Virgo collaboration for the search of continuous gravitational waves. In addition, interstellar scintillation can be studied with the same kind of data used at different time scales. Also, transient phenomena, such as fast radio bursts (FRBs), magnetars, and pulsar glitches are recent additions to the goals of our observations. Lastly, we discuss in Section~\ref{sec:conclusions} the impact that our contributions to radio observations in the Southern hemisphere may have on those areas of current astrophysical research, and the potential for near-term improvement in both hardware and software.
   

\section{The IAR observatory}
\label{sec:IAR}

\begin{figure}[ht]
  \centering
    \includegraphics[width=0.99\linewidth]{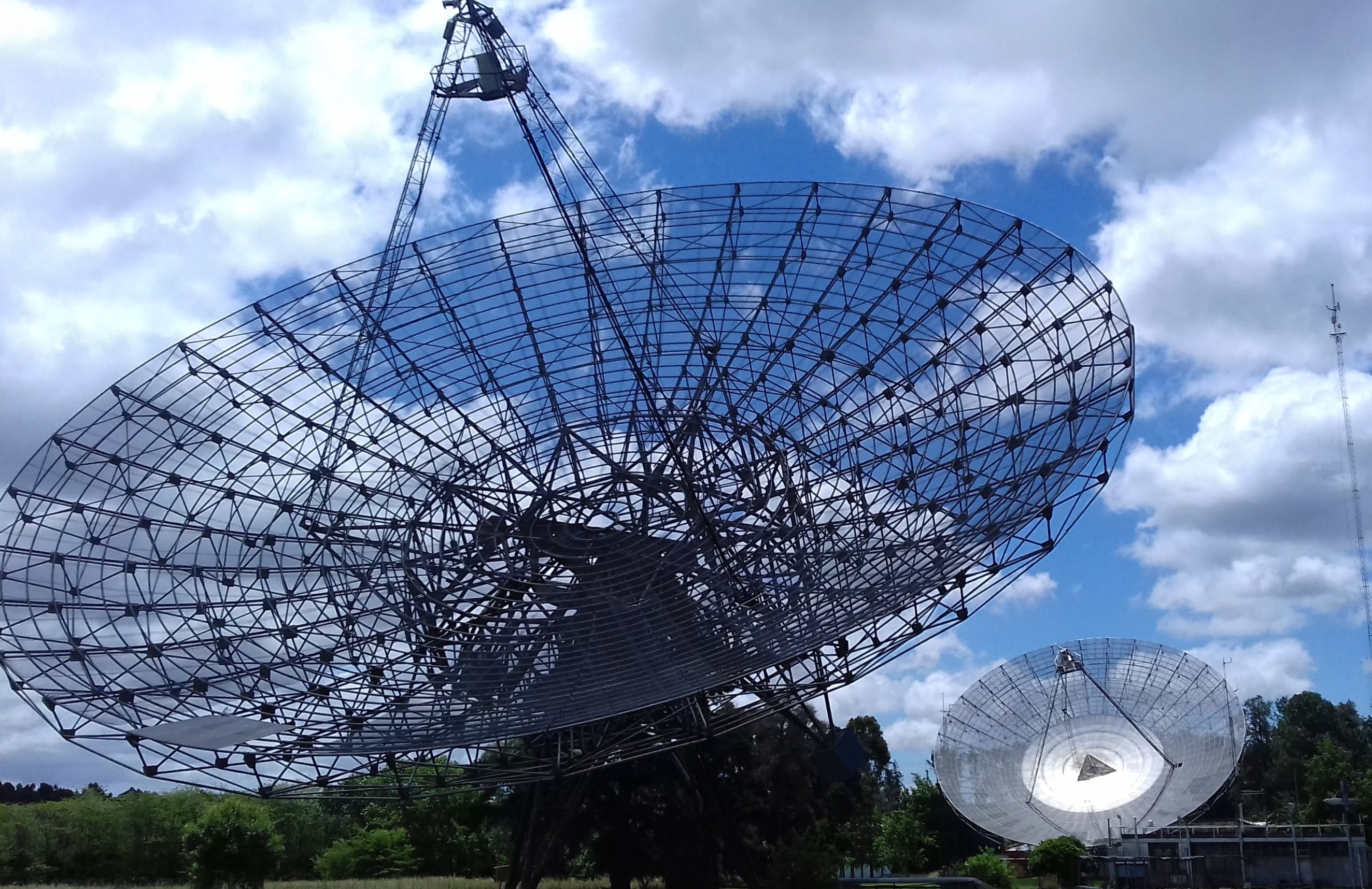}
    \caption{View of IAR antennas, A2 (left) and A1 (right).}
\label{fig:AntennaII&I}
\end{figure}

\begin{figure*}[ht]
  \centering
    \includegraphics[width=0.84\textwidth]{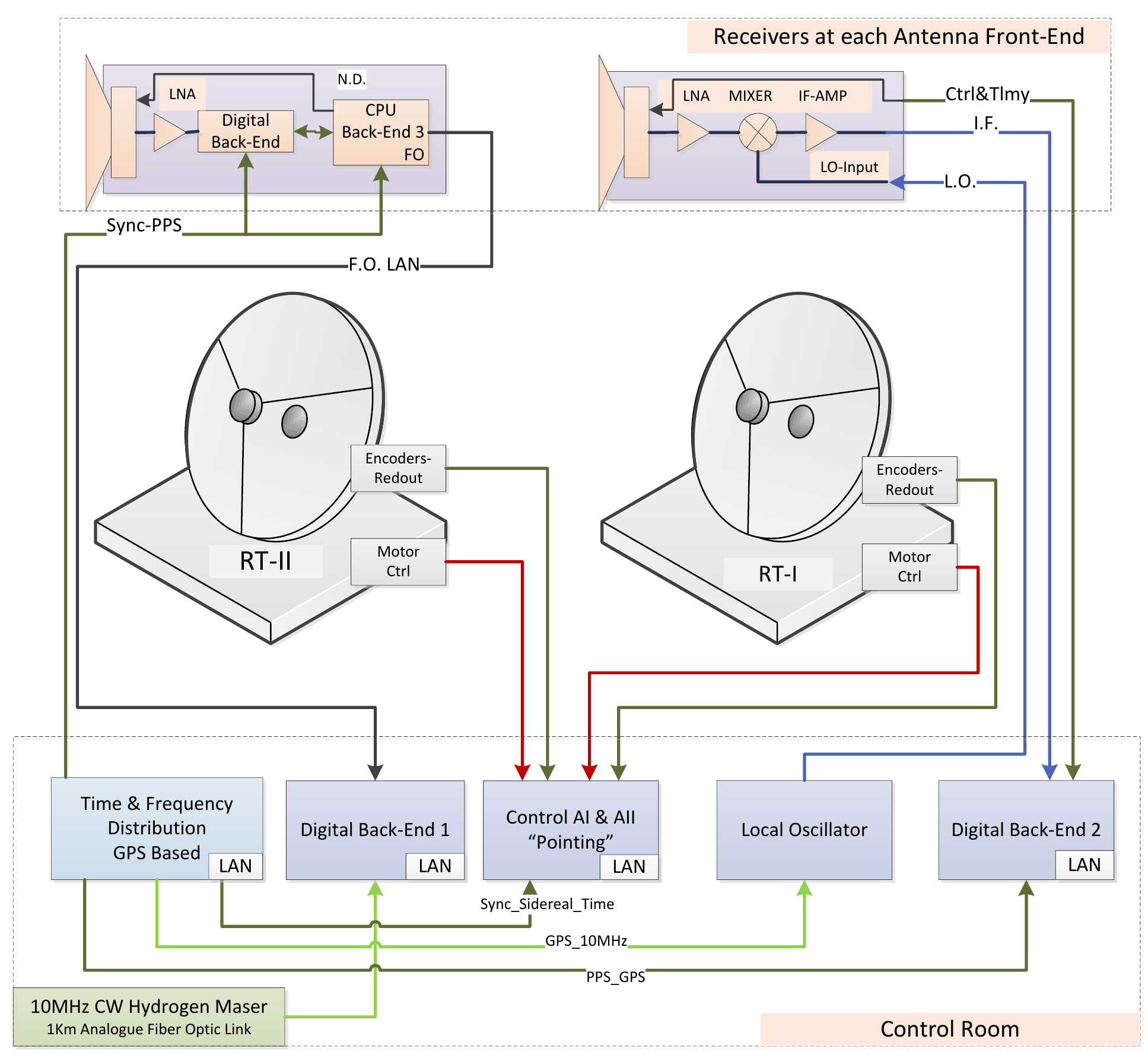}
    \caption{Current setup of IAR antennas.}
\label{fig:ReceiverBlocks}
\end{figure*}

Located in the provincial park Pereyra Iraola near the city of La Plata, Buenos Aires, the IAR itself is located at $-34\degr 51\arcmin 57\arcsec.35$ (latitude) and $58\degr 08\arcmin 25\arcsec.04$  (longitude), with local time UTC-3. The IAR observatory has two 30~meters diameter single-dish antennas, A1 and A2, aligned on a North-South direction (Fig.~\ref{fig:AntennaII&I}), separated by $120$~meters. These radio telescopes cover a declination range of $-90\degr < \delta <-10\degr$ and an hour angle range of two hours east/west, $-2\mathrm{h} <t< 2 \mathrm{h}$. The angular resolution at 1420 MHz is $\sim 30\arcmin$. The block diagram of Figure~\ref{fig:ReceiverBlocks} represents the connections of the antennas with their different modules.  

\subsection{Antenna 1}
Since 2004, several updates and repairs were made on A1, including a complete front-end repair in 2009 and a new set of positional encoders installed in 2014 to keep the tracking system up to date. In 2015, we installed a software defined radio (SDR) module to perform pulsar observations. The characteristics of the current frontend of A1 are listed in Table~\ref{table:antenna-param}.

The frontend has as a feeder system a orthomode transducer (OMT) with a quadridge type wave guide to coax transducer, using a $90\degr$ hybrid coupler it gives both circular polarization products (LHC and RHC) from the linear products. Currently only one circular polarization is used. We have inserted new band pass radio frequency (RF) filters into the receivers of both antennas; the rest of the analogue chain corresponds to a standard heterodyne receiver. A1 benefits from a lower insertion loss filter than A2 in the range 1150--1450~MHz, model ZX75BP-1280+ from Mini-circuits\footnote{\url{https://www.minicircuits.com/WebStore/dashboard.html?model=ZX75BP-1280-S\%2B}}.

The backend or acquisition module is based on two SDRs model B205 from Ettus\footnote{\url{https://www.ettus.com/all-products/usrp-b200mini-i-2/}} using a Xilinx Spartan-6 XC6SLX75 FPGA. 
This allows us to acquire raw samples from the front-end intermediate frequency as voltages in time series with a maximum rate of 56~MHz per board, and a universal serial bus (USB) 3.0 for connectivity. The sample rate is the same as the analogue bandwidth due to the internal frontend  in the SDR module. 
As each receiver has two digitizer boards, each with 56~MHz of bandwidth; we can use them in two modes: i) as consecutive bands, giving a total of 112~MHz of bandwidth for a single polarization; ii) by adding the two polarizations of 56~MHz bandwidth in order to obtain total power. A1 currently uses the first configuration.

The surface of A1 consists of a solid area at the center of the parabolic surface, while the rest is made of perforated aluminum sheets. This configuration gives an aperture efficiency of 32.8\% \citep{Testori:2001vp}.

\begin{figure}
  \centering
  \includegraphics[width=0.44\textwidth]{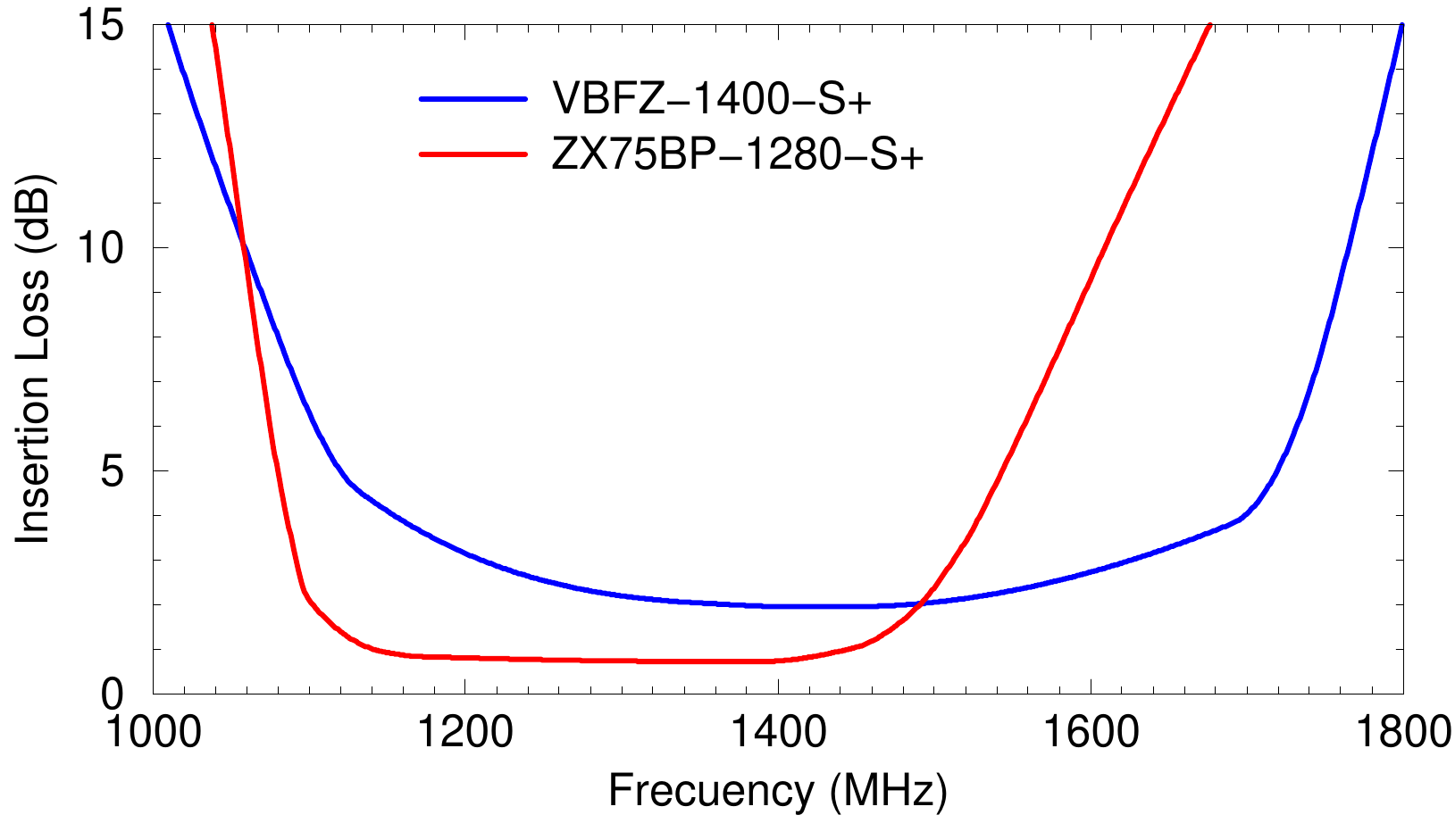}
    \caption{RF filters installed in A1 (ZX75BP-1280-S+) and A2 (VBFZ-1400-S+).}
\label{fig:filters}
\end{figure}

\subsection{Antenna 2}
A2 has a newly developed receiver that is fully operational since November 2018. The receiver, different from the one in A1, has the digitization stage directly in the RF band, having as a result less RF components and more RF bandwidth available at lower cost. The RF feeder uses a turnstile feed with two orthogonal circular polarization outputs, each output is connected to the SDRs through their low noise amplifiers and filters.

The backend of A2 has the same scheme as A1 but with a different configuration. For A2, both SDR boards take data of each circular polarization at the same time, frequency and bandwidth. The processing software adds both polarizations to obtain total power. At the present time, we are not processing the polarization products as Stokes parameters, though this will be implemented in the processing software shortly.

A2 has a wider range of sensitivity, including higher frequencies up to 1600~MHz, with the use of a filter model VBFZ-1400-S+, also from Mini-circuits\footnote{\url{https://www.minicircuits.com/WebStore/dashboard.html?model=VBFZ-1400-S\%2B}} (Fig.~\ref{fig:filters}). The characteristics of the current frontend in A2 are listed in Table~\ref{table:antenna-param}.

The surface of A2 presents a grid mesh for all its surface, giving a worse figure for its aperture efficiency, therefore resulting in a different Gain; the estimated aperture efficiency for A2 is 30.0\%.

\begin{table}
\caption{Parameters of the antennas and their receivers (frontend and software configuration).}
\centering
\begin{tabular}{lcc}
\hline \hline
Parameter                   &     A1        &   A2 \\ 
\hline 
Antenna diameter            & \multicolumn{2}{c}{30~m} \\ 
FWHM at 1420~MHz  & \multicolumn{2}{c}{$30\arcmin$} \\
Mounting                    & \multicolumn{2}{c}{Equatorial} \\
Maximum tracking time       & \multicolumn{2}{c}{220 min} \\
\hline
Low Noise Amplifiers\tablefootmark{a}   & HEMT He  & E-PHEMT \\
Filters Range (MHz)       & 1100$-$1510   & 1200$-$1600  \\
Electronics Bandwidth   & 110~MHz      & 200~MHz \\
Polarization                & One circular & Two Circular\\
Receiver Temperature        & 100~K             & 110~K \\
Aperture efficiency\tablefootmark{b} & 32.8\%     &   30\%   \\
Gain\tablefootmark{b} (Jy~K$^{-1}$) & 11.9  & 13.02 \\
Calibration                 & \multicolumn{2}{c}{Noise injection at feed} \\
\hline 
Instantaneous Bandwidth     & 112~MHz   & 56~MHz  \\
Polarization product        & One (circular)   & Total power  \\
SDR Models                  & \multicolumn{2}{c}{B210 - B205-mini-i} \\
Boards per CPU    & \multicolumn{2}{c}{Two} \\
Max data rate per           & \multicolumn{2}{c}{54~KHz} \\
Reference input             & \multicolumn{2}{c}{PPS} \\
Computer                    & \multicolumn{2}{c}{CPU i7, NVMe 1.2} \\
                            & \multicolumn{2}{c}{PCIe Gen 3x2 SDD} \\
Software language           & \multicolumn{2}{c}{C} \\
\hline
\end{tabular} 
\label{table:antenna-param}
\tablefoot{\tablefoottext{a}{At room temperature.} \tablefoottext{b}{Values from \cite{Testori:2001vp}.} }
\end{table}

\subsection{Clock synchronization}
Clock synchronization of the digital boards is performed with the one pulse per second (PPS) signal of a global positioning system (GPS) disciplined oscillator with an accuracy of $1.16 \times 10^{-12}$ (one-day average). In order to get a precise data time stamp, the PPS signal is used in the SDR boards to synchronize the first time sample with the \textit{exact} second of the GPS time. Then, the acquisition software reads the computer clock, which is synchronized with a Network Time Protocol (NTP) and a PPS signal from the GPS directly connected to the kernel OS through a serial port.

At less than one kilometer away from the IAR the Argentine-German Geodetic Observatory is located, AGGO\footnote{\url{http://www.aggo-conicet.gob.ar}}, dedicated, among other research, to Very Large Baseline Interferometry 
observations of quasars for geodetic purposes. This kind of measurements requires a precise clock for the synchronization with other observatories. This is achieved using an hydrogen maser clock with a short time stability of $10^{-15}$ (Allan Variance). With a locally developed RF-Over-Fiber device \citep{2013JInst...810003M}, we receive the 10~MHz signal from the AGGO's hydrogen maser using a fiber optic cable that connects both institutions. This signal is used to synchronize a test unit backend with the aim to compare results from the different time bases used at the moment. In the future, this shall be used to synchronize the data acquisition. 

\subsection{RFI environment}

\begin{figure}
  \centering
    \includegraphics[ angle=270, width=0.99\linewidth]{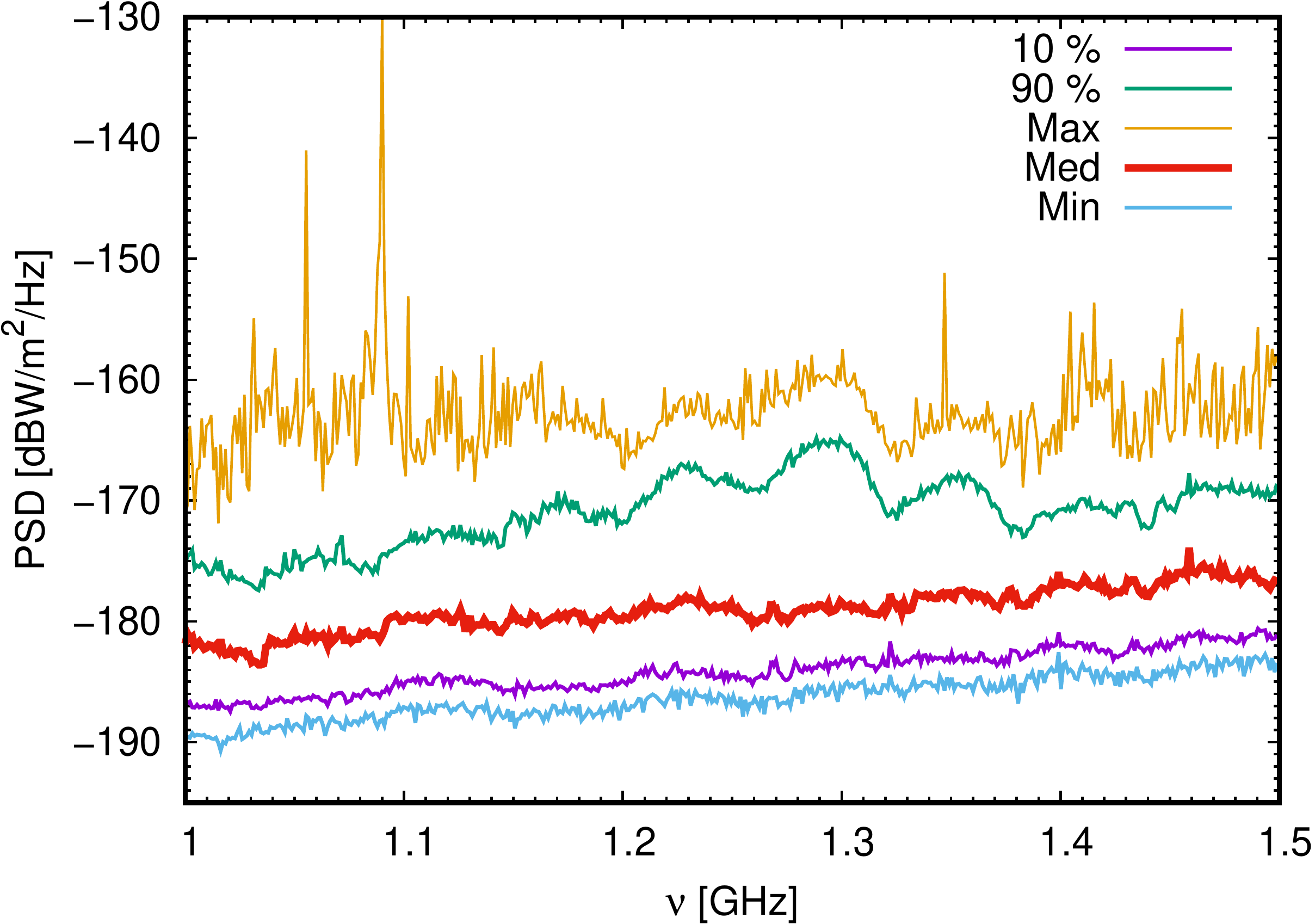}
    \caption{Wide bandwidth Power Spectrum Density obtained from a one-month average of measurements to characterize the RFI environment at IAR.}
\label{fig:rfia1}
\end{figure}

The IAR is located in a rural area outside La Plata, Buenos Aires. Although this is not a radio frequency interferences (RFI) quiet zone, the radio band from 1~GHz to 2~GHz has a low level of RFI activity that ensures the capability to do radio astronomy\footnote{Argentina is a  member of the International Telecommunication Union that protects radio bands for astronomical observations.}, as confirmed by the latest RFI measurement campaign from December 2017 \citep{2014ivs..conf..210G} for over a month in which the 90 \% of RFIs are detected below the $-$160~power spectral density (PSD) [dBW~m$^{-2}$~Hz$^{-1}$] (Fig.~\ref{fig:rfia1}). Moreover, the IAR has implemented a new protocol to provide a clean RFI local environment that is compatible with the current research and technical activity of the nearby antennas.
%

Observations with A1 have shown that RFIs affect $\sim 10\%$ of the daily observing time, although during night-time the RFI activity reduces significantly and compromises less than $1\%$ of the observing time. In average, observations with A2 have a significant less amount of RFI due to its distance to the administrative offices and laboratories. 

\begin{figure}[ht]
  \centering
    \includegraphics[width=0.95\linewidth]{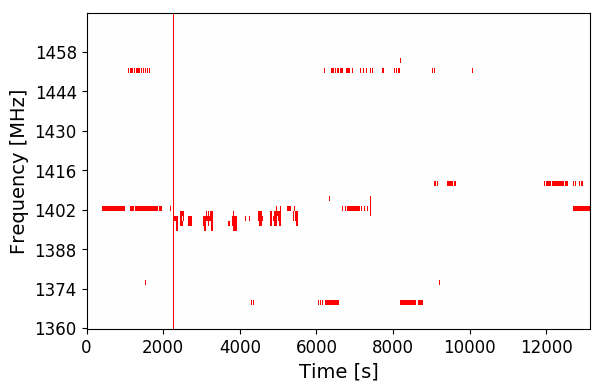}
     \includegraphics[width=0.95\linewidth]{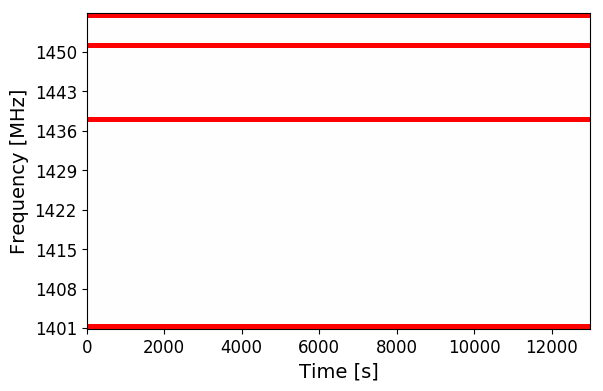}
    \caption{Example of a 3.5~hours RFI mask for A1 (top) and A2 (bottom) for a simultaneous observation using \texttt{rfifind} with \texttt{-time 1} and default parameters; coloured sections are masked out. The mask criteria acts differently in each antenna given their different sensibility, so the plots are not in the same scale.
    RFIs affect only $1.3\%$ of this observation for A1 and less than $6.3\%$ for A2. We note that the persistent RFIs in A2 are monochromatic and affect individual channels only (2 of 64 in this case). The frequency channels on the borders of the bandwidth in A2 are removed due to the design of the receiver.}
\label{fig:rfia1a2}
\end{figure}

Narrow-band mitigation is performed with the {\tt rfifind} package from PRESTO\footnote{\url{https://www.cv.nrao.edu/~sransom/presto/}}. First, the software parses the data in pieces of a certain time width (1~s in our case) per frequency interval. Then, it identifies in each one of these pieces weather the total power is too high, the data have an abnormal standard deviation, or the average of the data is above some given threshold. In that case, a mask is applied to the data before it is processed. In Figure~\ref{fig:rfia1a2} we present an example mask for A1 and A2 that shows the flagged data as a function of time and frequency channel. We note that A1 is more affected by random RFIs due to its local environment. These RFIs are usually mitigated during night time, out of office hours. In the case of A2, the RFIs are predominantly monochromatic and their impact can be mitigated by using a larger number of frequency channels. We also note that these RFIs were proven to be polarized, so it is not straightforward to compare the masks from A1 and A2. Moreover, the mask criteria applies differently in each antenna given their different sensibility. A more detailed analysis of this RFI environment is ongoing.

\subsection{Future upgrades}\label{sec:upgrades}

We plan to increase the bandwidth of the receivers. This implies: i) an upgrade of the frontend  to be able to operate in a frequency band from 1~GHz up to 2~GHz with insertion losses below 1~dB; ii) a new backend based on the CASPER boards like the SNAP board\footnote{\url{https://github.com/casper-astro/casper-hardware/blob/master/FPGA_Hosts/SNAP/README.md}} with bandwidths up to 500~MHz for each polarization. Moreover, the new receiver will benefit from state-of-the-art low-noise-amplifiers and electronics that will allow to reduce the system temperature to $T_\mathrm{sys} < 50$~K (using its cryogenic capability).

%
\section{Observations}\label{sec:obs}
%

\subsection{Software infrastructure}

\begin{figure}[ht]
  \centering
    \includegraphics[width=0.99\linewidth]{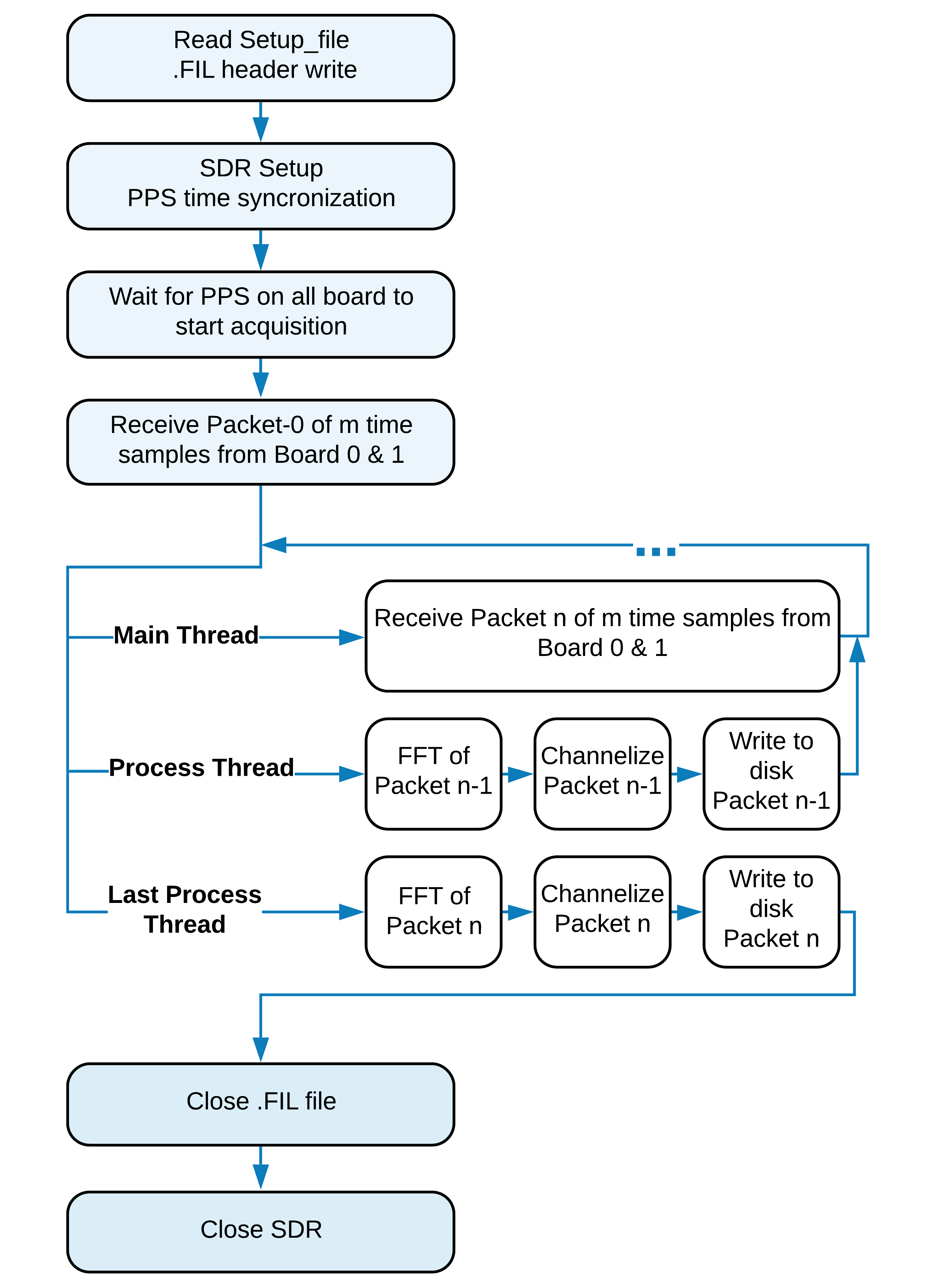}
    \caption{Pulsar software data acquisition block.}
\label{fig:swblock}
\end{figure}

The acquisition software was developed entirely at the IAR in \texttt{C} language. It processes the raw voltage samples at the desired rate without losses, while transforming 
them into a time series of RF channels. The software uses a scheme of 
synchronized threads in order to read the time samples from the different boards, process the fast Fourier transform (FFT) products, do the time average, and separate the final data product into channels before writing to disk. This is performed while keeping the synchronization with the PPS signal from the GPSDSO. The final product is a file in Filterbank (SIGPROC) format\footnote{\url{http://sigproc.sourceforge.net}}, compatible with standard pulsar reduction software like PRESTO \citep{2001AAS...19911903R, 2002AJ....124.1788R,2003ApJ...589..911R}. Figure~\ref{fig:swblock} represents the software diagram used in the \texttt{C} code, and Table~\ref{table:antenna-param} summarizes the main parameters of the digital receiver and the configuration used on each antenna.

Tracking and pointing systems of A1 and A2 are controlled remotely through the IAR server. A weather station and a video-camera help to control the IAR environment. Data acquisition is performed by a different computer, connected to the backends. After acquisition, Filterbank files are saved and processed in the IAR storage system. The raw data files are also transferred to a data center at the Rochester Institute for Technology (RIT-PuMA-DEN Lab) for backup. 

\subsection{Automation}

We are developing a distributed software architecture to control both IAR radio telescopes. Our goal is to generate a modern, dynamic and heterogeneous system in which modularity is an essential part, both in the development and in the expansion of the tools available for the observatory. 

\begin{figure}[ht]
  \centering
    \includegraphics[width=1.0\linewidth]{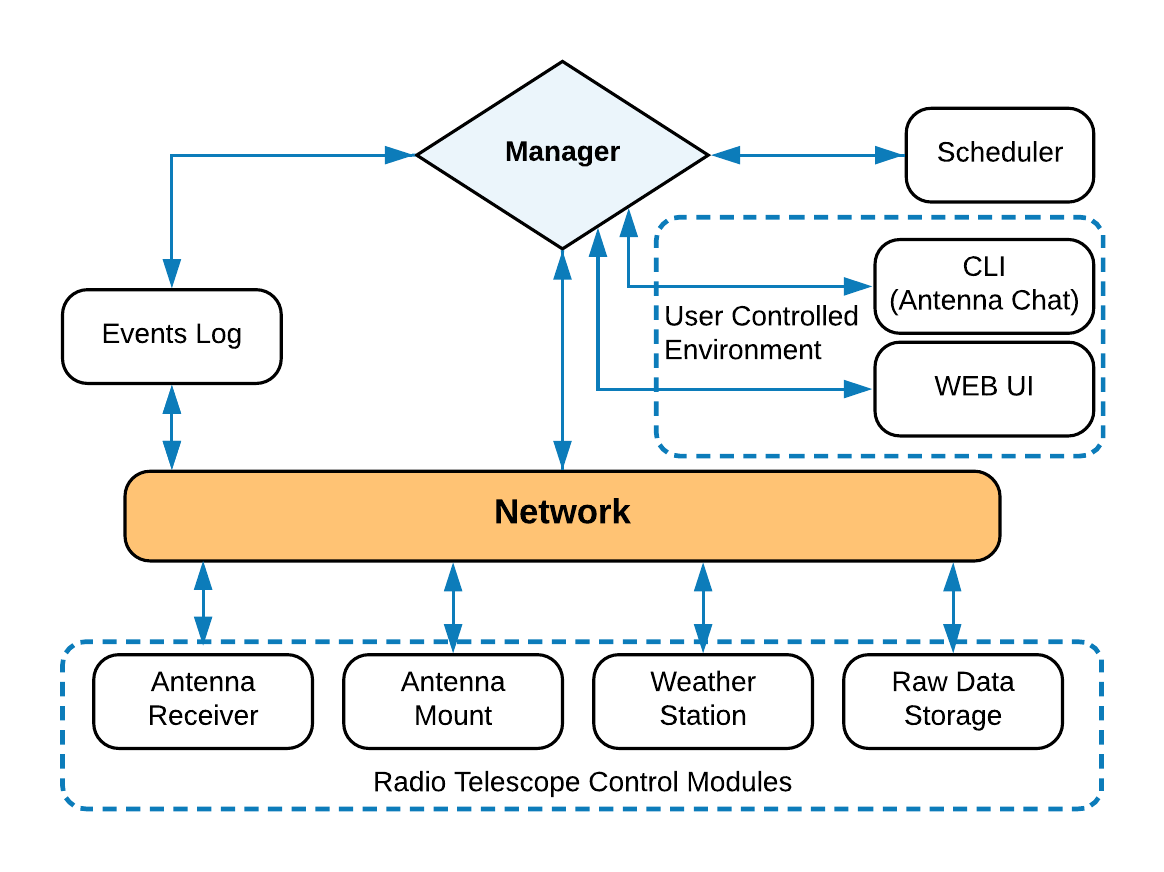}
    \caption{Software architecture for the control of the radio telescopes.}
\label{fig:radiotcontrol}
\end{figure}

We are working in upgrading the current client-server architecture by means of building a scalable and dynamic control software that consists on a series of simple modules that perform specific tasks. These modules can be orchestrated by states, events and messages passed to a controller software that has enough privileges to make decisions upon the running modules. This common communication interface/API allows the use of GUIs, CLIs and simple viewers. A scheme of the software architecture is shown in Fig.~\ref{fig:radiotcontrol}.

Finally, we plan to develop a scheduler to fully automate observations in order to offer the whole observation pipeline to the scientific community. In addition, the IAR is preparing a public proposer's interface together with online tools to assess the technical aspects of a requested observation and a remote monitoring during its performance.


\subsection{Data processing}

As we mentioned before, we apply PRESTO to process the Filterbank files acquired. First, we employ the \texttt{rfifind} routine to generate a mask, which allows us to remove the RFIs. Then, using that mask, we fold the data with \texttt{prepfold}. For this step, we use the ATNF catalogue\footnote{\url{http://www.atnf.csiro.au/research/pulsar/psrcat/}},][]{Manchester:2004bp} data as input. The outputs of the last routine are a set of \texttt{prepfold} (.PFD) files.

For the moment we are working on three different projects: detection of glitches, pulse time of arrival (ToA) extraction, and flux density measurement. In the case of a glitch search, we analyze if the pulsar observed period matches with the expected topocentric period for the observing date derived from the latest reported ephemeris in the ATNF catalog.

\begin{figure}[ht]
  \centering
    \includegraphics[width=0.9\linewidth]{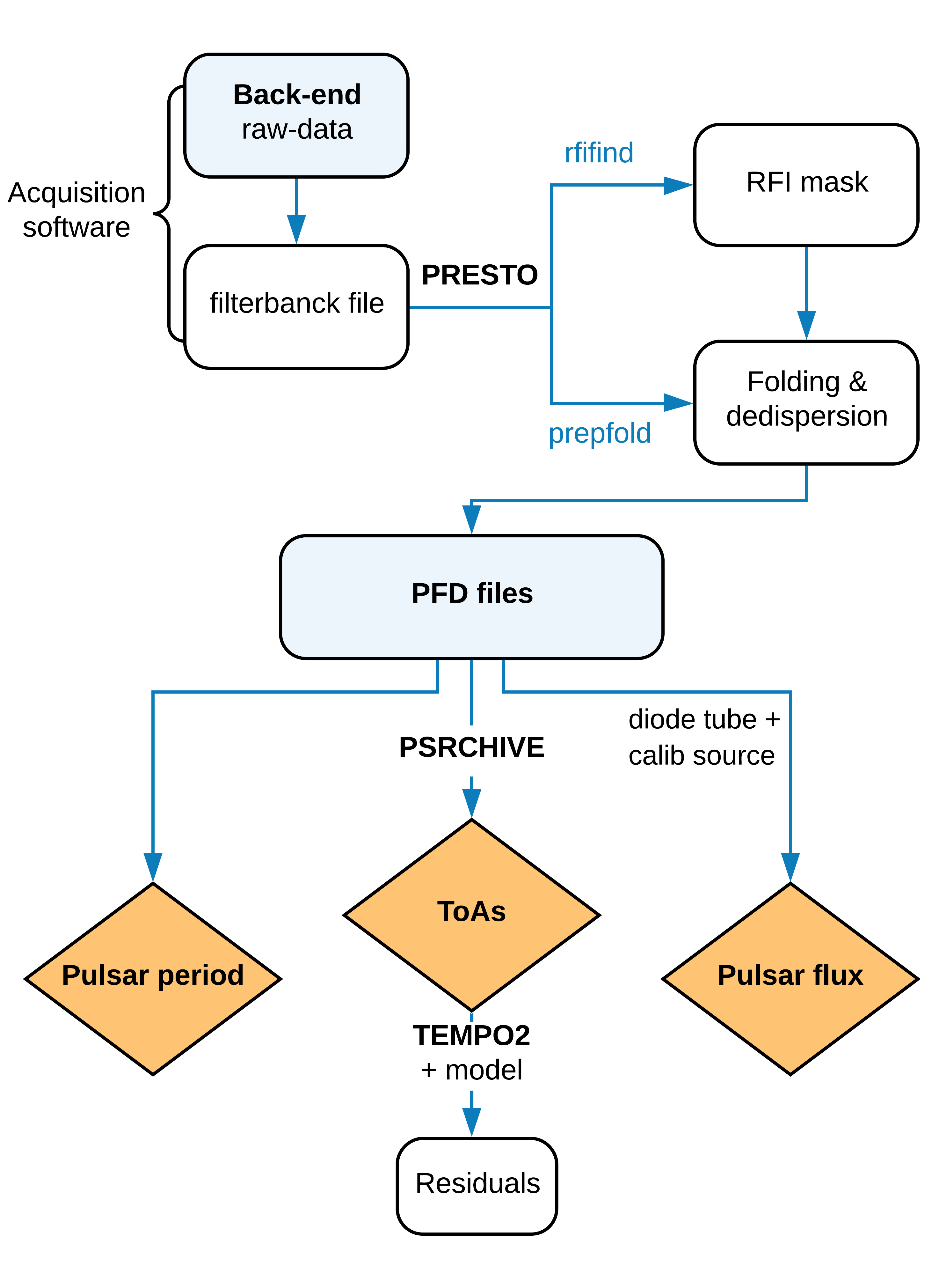}
    \caption{Data processing summary.}
\label{fig:datapipe}
\end{figure}

If it is not the case, we reprocess the Filterbank file doing a fitting for the new period (see Section \ref{sec:glitches}). The extraction of ToAs is handled with the PSRCHIVE \citep{Hotan:2004tz} package \texttt{pat}. These ToAs are processed with TEMPO2, using a suitable template for the pulse profile, to compute residuals needed for our scientific goals (see Section \ref{sec:pulsartiming}). The third project is the calibration of pulsar flux densities. For that, we employ the diode tube in each of our radio telescopes and calibration sources, such as Hydra~A. Figure~\ref{fig:datapipe} shows a summary of the whole reduction process of the data.

Further post-processing is carried out with the freely available PyPulse package\footnote{https://github.com/mtlam/PyPulse}. For instance, we make use of this package to compute the signal-to-noise ratio of our observations defined as the ratio between the mean pulse peak and the RMS of the noise. The latter is calculated from an off-pulse window of size $1/8$ of the total phase bins in which the integrated flux density is minimal. We refer to \cite{Lam:2016} for further details.

\subsection{Observational capabilities and testing} \label{sec:observational_capabilities}

We have been observing with both antennas since December 2018 to test and calibrate them. Both radio telescopes are capable of observing sources for almost four hours on a daily basis. We note that bright pulsars with sharp-peaked profiles are the easiest to detect; this is straightforward from the standard formula for the expected signal-to-noise ratio \citep[S/N;][]{lorimer2012handbook}:
\begin{equation} \label{eq:snr}
    \mathrm{S/N} \lesssim S_\mathrm{mean}  \frac{\sqrt{n_\mathrm{p}  t_\mathrm{obs} B}}{G T_\mathrm{sys}} \sqrt{\frac{P-W}{W}},
\end{equation}
where $S_\mathrm{mean}$, $P$, and $W$ are the mean flux density, period and equivalent width of the pulses, respectively, $G$, $B$, $n_\mathrm{p}$, and $T_\mathrm{sys}$ are the antenna gain, bandwidth, number of polarizations, and system temperature, respectively, and $t_\mathrm{obs}$ is the effective observing time. Using Eq.~(\ref{eq:snr}) we can estimate which pulsars can be observed with IAR antennas. We introduce the parameters given in Table~\ref{table:antenna-param} in the equation and we set the maximum observing time to $t = 220$~min, from where we obtain a value for the expected S/N. Considering that a reliable detection can be achieved for those pulsars with a $\mathrm{S/N} > 10$, we make use of the {\tt python} tool \texttt{psrqpy} \citep{psrqpy} to select the ones that IAR's antennas are capable to detect\footnote{Note that this estimated way to calculate the signal-to-noise correspond to the S/N as computed with a cross-correlation function instead of the peak to off-peak rms S/B, see \cite{Lam:2016} to the mathematical relation between the two quantities.}.

\begin{figure}[ht]
  \centering
    \includegraphics[width=0.9\linewidth]{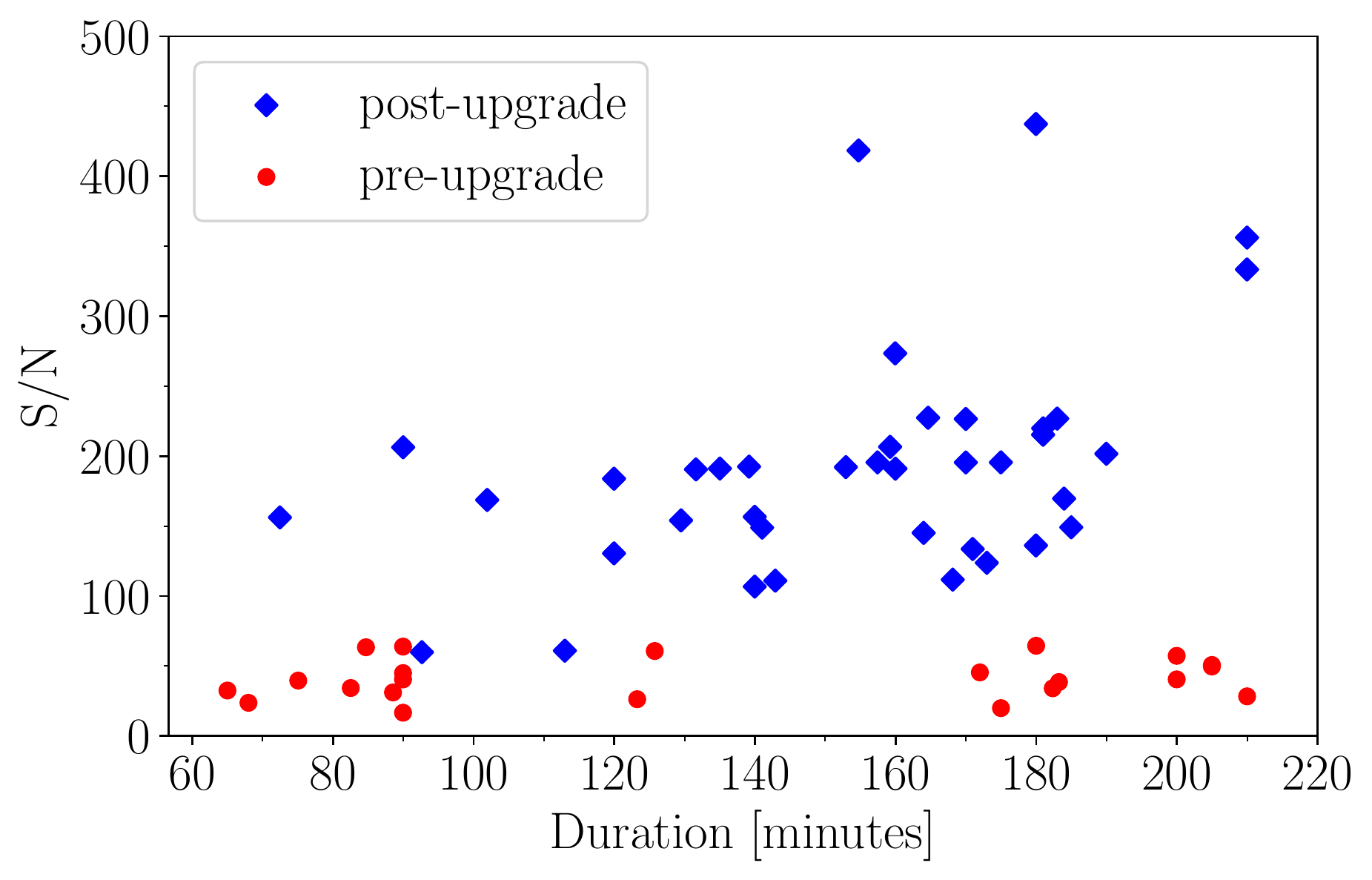}\\
    \includegraphics[width=1.\linewidth]{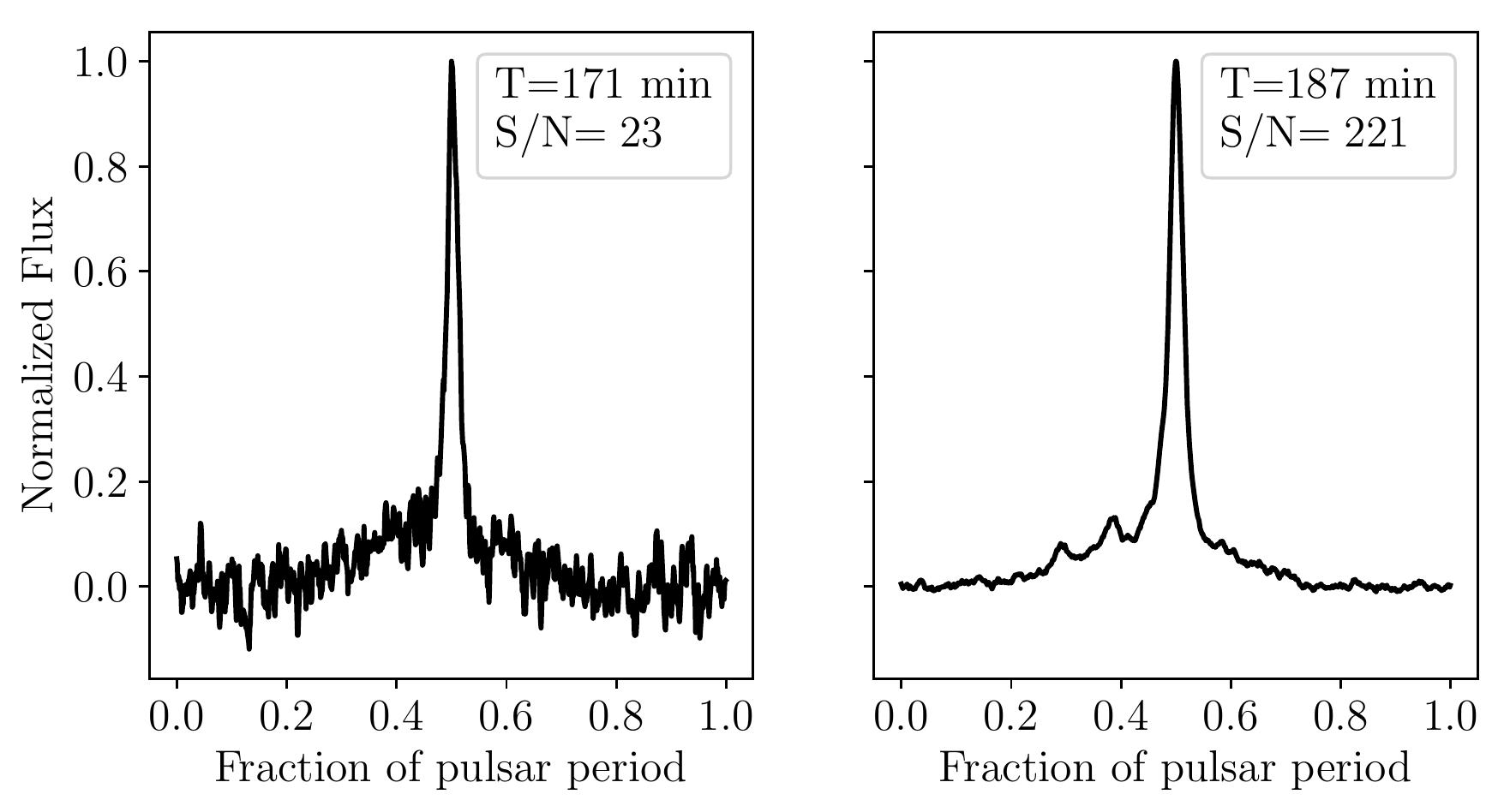}
    \caption{Top panel: Signal-to-noise distribution as a function of the integration time for different observations of J0437$-$4715 before and after the upgrade in A1. Bottom panel: typical pulse profiles in the pre-upgrade and post-upgrade configuration.}
\label{fig:J04-profile}
\end{figure}

Our main target for testing purposes is the bright MSP J0437$-$4715 (Fig.~\ref{fig:J04-profile}). This pulsar allows us to test the timing quality of both antennas thanks to its high timing stability, high brightness, and short spin period. In Fig.~\ref{fig:J04-profile}, we compare our observations of J0437$-$4715 in the pre-upgrade and post-upgrade configuration. The upgrade consisted in an increase in bandwidth from 20 to 112~MHz, the incorporation of a better band-pass filter, and the use of both digitizer boards. 

We note that this pulsar shows an important variation of the flux due to scintillation \citep{2014MNRAS.441.3148O}, most notably in observations after the upgrade which have a larger bandwidth. We show the preliminary timing results in the next section.

\subsection{Collaborations with other radio observatories in Argentina} \label{sec:extended_observations}

One spin-off from the development of the digital back-end receiver is its use as a stand alone unit for radio astronomical observations. This unit has been tested successfully in a 35-meter Deep Space Antenna (DSA) named CLTC-CONAE-Neuquen
\footnote{\url{https://www.argentina.gob.ar/ciencia/conae/centros-y-estaciones/estacion-cltc-conae-neuquen}}, installed in Neuquen, Argentina. Test observations were carried out in S and X bands targeting both pulsars (Vela and J0437$-$4715) and continuum sources. This will enable the possibility to do simultaneous observations between the IAR radio telescopes and the DSA antenna at different frequencies for radio astronomical research.

%
%
\section{Enabling science projects}\label{sec:science}
%

In this section we describe several scientific projects that we are ---or will be--- able to perform with the refurbished IAR antennas. We present the current state of each project, some preliminary results, and projections with future hardware improvements.

\subsection{Pulsar timing and gravitational waves}
\label{sec:pulsartiming}
Millisecond pulsars show a remarkable rotational stability. This allows one to predict the TOA of their pulses with high precision over long periods of time. Given a physical model, we can compare the predicted TOAs, $t_\mathrm{P}$, with the actual observed TOAs in a certain reference frame, $t_\mathrm{O}$, and compute the residuals $\delta t = t_\mathrm{P} - t_\mathrm{O}$. 
The residuals contain information of the astrophysical system and small effects due to different processes that can be incorporated in the timing model. 
For millisecond pulsars in particular, the residuals are dominated by white-noise when a large-enough bandwidth is used to resolve accurately the dispersion measure and can reach the order $\delta t<1\mu$s.

With this idea, pulsar timing arrays (PTAs), consisting usually of tens of precisely-timed MSPs, can be used as a Galactic scale detector of low-frequency gravitational waves. The main goal is to detect a ``stochastic background'' of such low-frequency gravitational waves, originated from an ensemble of unresolved supermassive black hole binaries (SMBHBs). Specifically, the effect of this background would appear in the PTA as a particular spatial angular correlation of the ToAs from different pulsars given by the Hellings-Downs curve \citep{Hellings:1983fr}. Several physical effects need to be modeled and corrected in order to find the effect of gravitational waves on the ToAs; chiefly, the pulsar dynamics and intrinsic instabilities, timing delays due to the interstellar medium and solar wind \citep[see][for a recent review]{Hobbs:2017zve}. There are three main PTA collaborations, NANOGrav from North-America \citep{Arzoumanian:2017puf}, EPTA from Europe \citep{Desvignes:2016yex}, and PPTA from Australia \citep{Reardon:2015kba}, together with an international PTA consortium \citep[IPTA;][]{Verbiest:2016vem} that coordinates common efforts. Observatories in China, South-Africa, India, and Argentina plan to join the IPTA soon.


Bright MSPs, such as PSR J0437$-$4715, are excellent targets for IAR's antennas. We are currently performing an almost four-hours per day monitoring of PSR J0437$-$4715 at 1400~MHz. These observations are projected to increase the sensitivity of pulsar timing arrays by increasing the observing cadence by a factor 20-30 and hence be sensitive to closer to merger (or less massive) SMBHB systems, and to reach potentially detectable SMBHB \citep{Zhu:2015tua} in their host galaxies \cite[e.g., Figure 2 in][]{Burt:2010jh}. These observations also provide an overall sky coverage together with Parkes and MeerKAT observatories on the Southern Hemisphere. Moreover, the almost 4 hours per day of data that IAR can provide of J0437$-$4715 allows to minimize statistical uncertainties due to jitter of the pulses, improving the timing quality \citep[see][]{shannon2014limitations,lam2018optimizing}. In our current set-up, the most limitating factor is the bandwidth. Full details of the IAR's contribution to J0437$-$4715 timing and future projections will be presented in an upcoming work. 

\begin{figure}
  \centering
    \includegraphics[angle=270, width=1.0\linewidth]{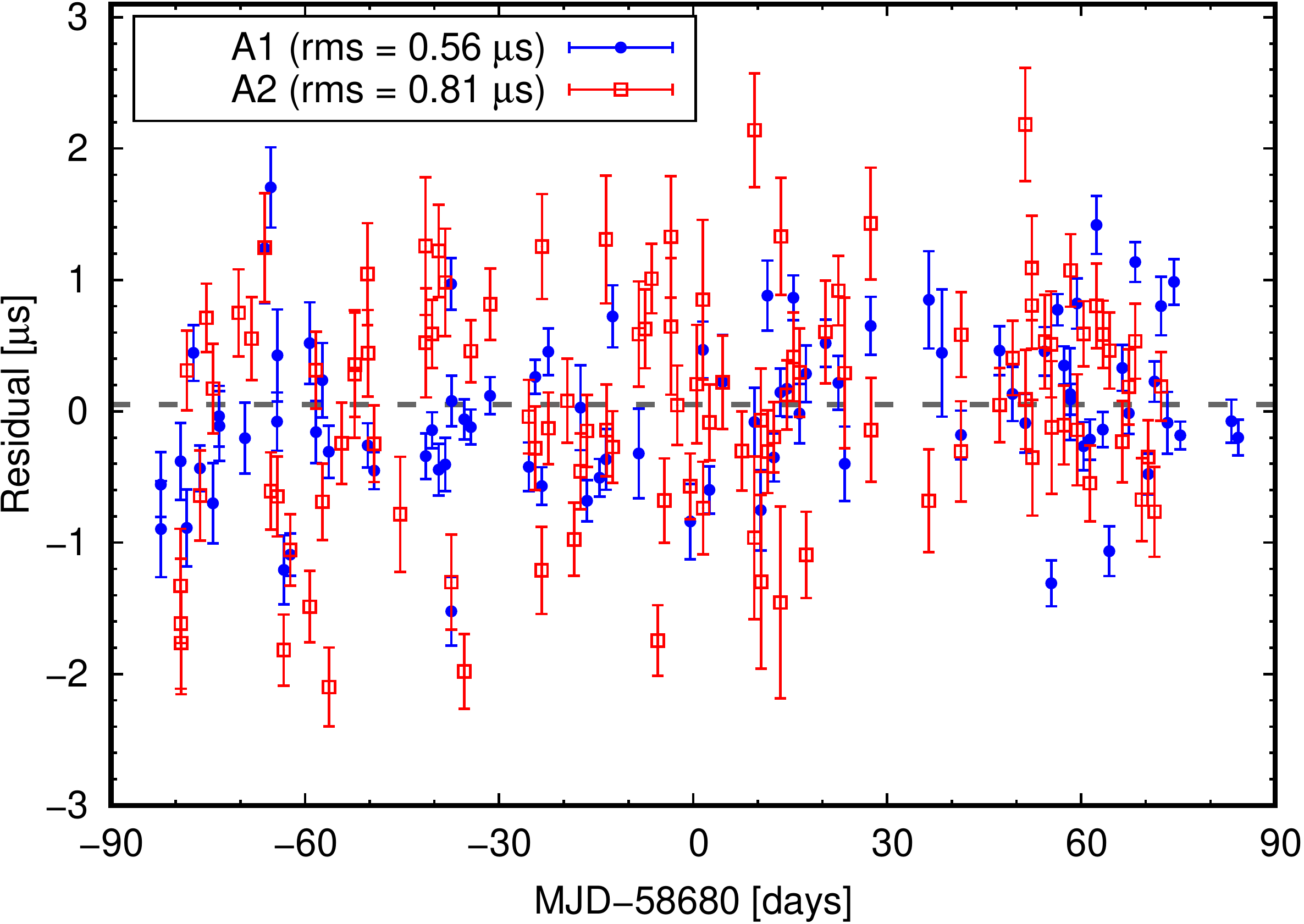}
    \caption{Residuals of the ToAs of the pulsar J0437$-$4715 measured at IAR with A1 
    and A2. 
    The residuals and $rms$ have been computed with the TEMPO2 package \citep{Hobbs:2006cd} and error bars correspond only to template-fitting errors (i.e., no systematics considered). }
\label{fig:ResidualsJ0437}
\end{figure}
Current observations of J0437$-$4715 at IAR lead to residuals root mean squares ($rms$) of $0.55~\mu$s for A1, $0.81~\mu$s for A2, and $0.78~\mu$s for A1+A2 (introducing a jump for matching both data sets), as displayed in Fig.~\ref{fig:ResidualsJ0437}. These values of $rms \lesssim 1~\mu$s are compatible with Table~1 of \cite{Burt:2010jh} expectations, but still far from PTA reported $0.1~\mu$s in \cite{Perera:2019sca} or the $0.04~\mu$s optimal reachable \citep[see][]{Oslowski2011}. The timing precision (i.e., RMS statistics or the single TOA precision) will improve with a larger bandwidth (up to 1~GHz). In addition, the precision of the timing parameters will improve with the continuation of daily observations to accumulate long term data (several years).
Other MSPs within the reach of IAR observatory are given in Table~\ref{tab:ms} \citep[extracted from the ATNF pulsar catalog\footnote{\url{http://www.atnf.csiro.au/research/pulsar/psrcat/}},][]{Manchester:2004bp}, detailing their barycentric periods, mean flux densities at 1400~MHz, binary models, dispersion measures, and numbers of glitches observed.

\begin{table}
\centering
\caption{Observable MSPs from IAR with the current setup (top) and with future upgrades (bottom). $P_0$ is the barycentric period of the pulsar, $S_{1400}$ the mean flux density at 1400~MHz, $W_{50}$ the pulse width at 50\% of peak, S/N the expected signal-to-noise with 220' of observation with A1, $D$ the distance based on the \cite{Yao_2017} electron density model, and $DM$ the dispersion measure.} 
\label{tab:ms}
\setlength{\tabcolsep}{4.5pt}
\begin{tabular}{lcccccc}
\hline \hline
JNAME           &  $P_0$     &  $S_{1400}$ & $W_{50}$\tablefootmark{a}  & S/N\tablefootmark{b}   & $D$ & $ DM $\\
                &  [ms]      &  [mJy]   & [ms]      &       &   [kpc]   &      [pc cm$^{-3}$] \\
\hline
    J0437$-$4715  & 5.757    &  150.2     & 0.141     & 336\tablefootmark{c}     &   0.16    &   2.64    \\
    J1744$-$1134  & 4.075    &  3.2    & 0.137     & 15.1     &   0.15    &   3.14    \\
    J2241$-$5236  & 2.187    &  1.95    & 0.07      & 9.4      &   0.96    &   11.41   \\
    J1643$-$1224  & 4.622    &  4.68    & 0.314     & 15.26     &   0.79    &   62.41   \\
    J1600$-$3053  & 3.598    &  2.44    & 0.094     & 13.1     &   2.53    &   52.33   \\
    J2124$-$3358  & 4.931    &  4.5    & 0.524     & 11.5      &   0.36    &   4.60    \\
    J1603$-$7202  & 14.84   &  3.5    & 1.206     & 10.4     &   1.13    &   38.05   \\
    J1730$-$2304  & 8.123    &  4.00    & 0.965     & 9.6       &   0.51    &    9.62   \\
    J0900$-$3144  & 11.11   &  3.00    & 0.8       & 9.5      &   0.38    &   75.71   \\
    J0711$-$6830  & 5.491    &  3.7    & 1.092     & 6.5      &   0.11    &   18.41   \\
    J1933$-$6211  & 3.543    &  2.30    & 0.36      & 6.0      &   0.65    &   11.52   \\
    J1652$-$48    & 3.785    &  2.70    & *         & *         &   4.39    &   187.8   \\
    \hline
\end{tabular}
\tablefoot{\tablefoottext{a}{Quoted values are indicative only, as the width of pulse at 50\% of peak is a function of both observing frequency and time resolution. \tablefoottext{b}{Estimated according to Eq.~(\ref{eq:snr}) and the values of A1 given in Table~\ref{table:antenna-param}. We note that Eq.~(\ref{eq:snr}) involves the equivalent width of the pulse $W$ which we do not know, but for a gaussian pulse it is valid to approximate $W \approx W_{50}$.} \tablefoottext{c}{The pulse of J0437$-$4715 significantly differs from gaussian, so in this case we use the equivalent width of the pulse that we measure, $W \approx 0.77$~ms.}}
} 
\end{table}

The detection of gravitational waves from compact binaries, along with their electromagnetic counterparts, notably enhances our comprehension of astrophysical processes. In particular, the case of the binary neutron star merger observed by LIGO-Virgo \citep{GBM:2017lvd} gave birth to full fledged multi-messenger astronomy. In connection with PTA detections of single sources, the modeling of accreting matter around merging SMBHBs 
and the characteristic features of their electromagnetic spectra is currently an extremely active research area
\citep{Bowen:2017oot,dAscoli:2018fjw,Bowen:2019ddu}. Likewise, the gravitational waves from merging supermassive black holes can present distinctive features in each polarization, which can inform us about the strong precession of the binary systems \citep{Lousto:2018dgd}. In addition to our contributions to PTAs collaboration, we plan to search (or at least place constraints) for continuous gravitational waves from individual SMBHBs \citep{Zhu:2015tua,Kelley:2017vox}. The cadence of daily rather than monthly observations leads to sensitivity to sources of gravitational waves being produced closer to the merger of the supermassive black holes by a factor $30^{2/3}$\citep[see e.g.][]{Blanchet:1996pi}, hence an order of magnitude increase in its amplitude $h$.
In particular, for sources of a few billion solar masses at $z=1$, we expect to reach Earth with gravitational strains oscillations of up to $h\sim10^{-14}$ \citep[see the case of QSO 3C 186 in][]{Lousto:2017uav}. Detection of gravitational waves from individual SMBHB \citep{Detweiler:1979wn,Hellings:1983fr,Zhu:2015tua} (as opposed to the stochastic background) may lead to important clues about the formation and evolution of such sources and can be performed by studying a few very well timed pulsars, like J0437$-$4715.

Another interesting millisecond pulsar to study with the next-generation backend is PSR J2241$-$5236 \citep{2011MNRAS.414.1292K}. 
This pulsar is currently observed by Parkes and MeerKAT and will be essential for the IPTA because of its excellent timing quality (D. Reardon, private communication). However, since it is a black widow pulsar with an orbital period of $\sim 3.5$~h, it shows orbital noise that reduces sensitivity to gravitational waves. However, such noise can be characterized and modeled with high-cadence observations. This makes this pulsar an excellent target for IAR as it can monitor the full orbit on a daily basis. A phase-resolved analysis is expected to be viable with the future improvements in bandwidth and sensitivity.

\subsection{Targeted pulsar studies for continuous gravitational waves detection from laser interferometry}


In addition to the most remarkable detection of merging binary black holes and binary neutron stars \citep{LIGOScientific:2018mvr}, the LIGO-Virgo collaboration monitors over 200 pulsars, looking for continuous gravitational waves coming from any time-varying (quadrupolar or higher) deformation of the spinning neutron stars \citep{Abbott:2017ylp,Authors:2019ztc}. The criteria to choose those pulsars is that their periods be less than 0.1~s, so the frequency of the emitted waves is at the start frequency of the LIGO-Virgo sensitivity curve, i.e. currently with a frequency above 20~Hz.
An important benchmark is given by the spin-down limit, obtained by equating the (radio) observed slowdown spinning rate to the expected rate due to the loss of energy by gravitational waves \citep{Palomba:2005na}.  
For the Crab (PSR J0534+2200) and Vela (PSR~J0835$-$4510) pulsars, current upper gravitational waves bounds show that this limit is now surpassed by nearly an order of magnitude, while for other six studied pulsars this spin-down limit was also recently reached \citep{Authors:2019ztc}.

The timing of those targeted pulsars is very important as their ToAs are used to construct ephemeris to search at specific frequencies in the LIGO-Virgo data. It is also important to know if and when those pulsars have glitches. In fact, overlooking glitches has a negative impact on standard CW analyses \citep{Ashton:2017wui,Ashton:2018qth}. While identifying glitches can help to better model \citep{Prix:2011qv} potential emitters of continuous gravitational waves 
\citep[see][for the 2006 Vela glitch]{Abadie:2010sf}.
During the O1/O2 LIGO observing runs, several of the following pulsars have displayed glitches, like PSR J0205+6449 during O1 and five others during O2, including Crab and Vela pulsars \citep{Keitel:2019zhb}; the others are PSR J1028$-$5819, PSR J1718$-$382, and PSR J0205+6449 \cite[see][]{Abbott:2017ylp,Authors:2019ztc}.


The third generation of laser interferometer detectors, with increased sensitivity in the low frequency band (starting at a few hertz), are potentially sensitive enough to hold a chance to observe continuous gravitational waves from selected MSPs or from younger glitching ones
\citep{Glampedakis:2017nqy}.

In particular, the pulsar J0711$-$6830 (at a distance of 0.11~kpc) is within a factor of 1.3 of the spin-down limit (assuming a $10^{38}$~kg~m$^2$ standard moment of inertia), and is one of our targeted pulsars for future observation at IAR. Another close-to-Earth recycled MSP, and close to its spin-down limit, is PSR J0437$-$4715, which is already being daily followed up at IAR. Other pulsars of interest for LIGO-Virgo \citep[Table~2 of ][]{Authors:2019ztc} that are on the reach of IAR's (future) observation capabilities (see Sect.~\ref{sec:observational_capabilities}) are pulsars J1744$-$1134, J1643$-$1224, J2241$-$5236, J2124$-$3358, J1603$-$7202, J0900$-$3144, and J1730$-$2304 (as displayed in Table~\ref{tab:ms}).


The criteria for pulsar selection for a direct detection of continuous gravitational waves is similar to that for pulsar timing arrays, since both require (preferable non-glitching) MSPs in order to extract signals from observations over years \citep{Woan:2018tey}. But while pulsar timing prefers more stable, recycled pulsars, young pulsars with larger asymmetries would be stronger gravitational waves emitters.
Pulsars located close to Earth are preferred both for direct gravitational waves, due to the larger amplitude of the waveform strain (inversely proportional to the Earth-pulsar distance), and for the pulsar timing arrays, in order to obtain a better signal (pulse profile) to noise (instrumental and interstellar media) ratio. These conditions suit well with IAR capabilities, with an added value of daily observations that allow for shorter time scales studies than most observatories.

\subsection{Magnetars}

Magnetars are isolated young neutron stars with very large magnetic fields (of the order of $10^{15}$~G). About 30 magnetars have been reported, but a much larger population is expected, given their transient nature. \textit{Fermi} and \textit{Swift} satellites observe them in soft gamma-rays and in X-rays associated with soft-gamma-ray repeaters (SGRs) and anomalous X-ray pulsars (AXPs).  These are characterized by energetic winds, intense radiation, and a decaying magnetic field on scales from days to months. For a recent overview of the observational (including radio) properties of magnetars we refer to \cite{Esposito:2018gvp} and, for its physical modeling, to \cite{Turolla:2015mwa}.

A possible connection with superluminous supernovae (SL-SNe) has been speculated. The explosions of these particular SNe are an order of magnitude more luminous than standard SNe and may lead to the formation of a highly-magnetized and fast-spinning magnetar, which in turn may energize the supernova remnants \citep{Inserra:2013ila}.
Magnetars may also be related to FRBs \citep{Eftekhari:2019jah}. LIGO has searched \citep{others:2016ifn} for coincident FRB and gravitational waves signals in its first generation runs (2007--2013), and for magnetar bursts during the advanced LIGO-Virgo observations \citep{Abbott:2019dxx}.

A few magnetars can be detected pulsating in radio wavelengths. They tend to display numerous glitches, and sometimes anti-glitches \citep[i.e. spin-down;][]{Archibald:2013kla}. Therefore, to study their early behavior, they need to be monitored with high cadence. Magnetars appear in the upper-right corner of the $P-\dot{P}$ diagram, i.e., they have large periods and period derivatives \citep{lorimer2012handbook}. Under the assumption that those pulsars only brake due to dipole radiation emission ($B \propto \sqrt{P \dot{P}}$), they show the highest magnetic fields known, which gives them their name. Their spectrum is roughly $S_\nu \propto \nu^{-0.5}$, thus harder than that of regular pulsars, which have $S_\nu \propto \nu^{-1.8}$ \citep[see for instance][]{Dai:2015awa}.

Another characteristic feature of magnetars is that their pulses do not stabilize in shape, as opposed to regular pulsars that reach stability after average of a few hundred pulses  \citep[although they might switch between a few of them on minutes to hours scales;][]{Esposito:2018gvp}. Also, magnetars can be very bright (reaching 10~Jy in the L-band) in radio and also have an on-off-on behavior \citep{Esposito:2018gvp}.

\begin{figure}
  \centering
    \includegraphics[width=1\linewidth]{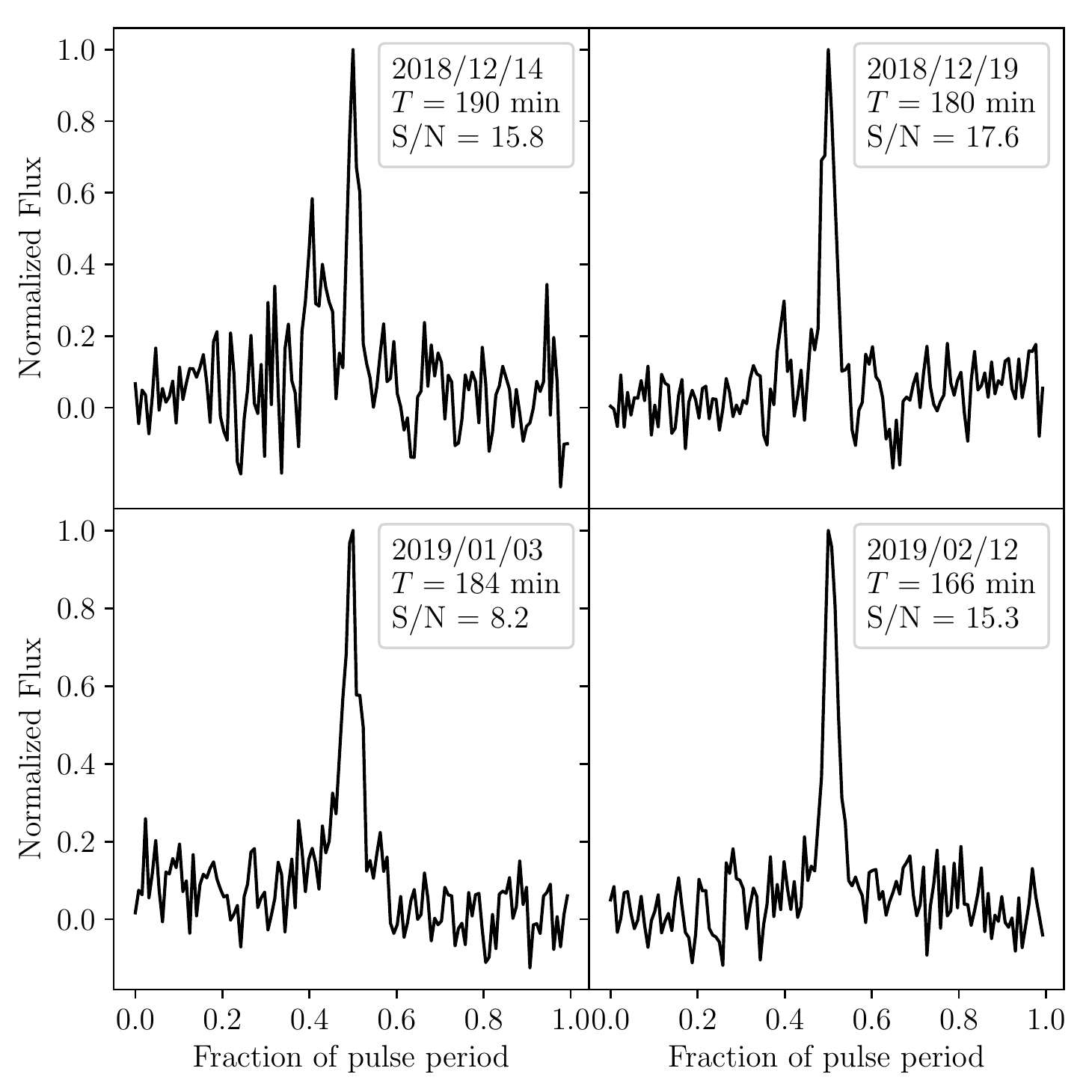}
    \caption{Pulse profiles of the magnetar J1810$-$197 at 1400~MHz measured by IAR antennas at different epochs.}
\label{fig:magnetar}
\end{figure}

The magnetar XTE J1810$-$197 has experienced periods of activity in X-rays \citep{Ibrahim:2003ev} and in radio frequencies, being the first magnetar in which radio pulsations were detected \citep{Camilo:2006zf}. After being in a radio-quiet state for several years \citep{Camilo:2016wfk}, this magnetar has recently experienced another outburst \citep{Lyne2018}. As an exploratory study we dedicated observing time to this object from Dec.~14, 2018 to Mar.~1, 2019 \citep{atel_magnetar}. Single-polarization observations with a bandwidth of 56~MHz centered at 1420~MHz revealed significant pulsating radio emission from XTE J1810$-$197 with a barycentric spin period of $P = 5.54137(3)$~s on MJD 58466.615, consistent with the values reported in \cite{Lyne2018}. Unfortunately, we could not derive polarization angles and calibrated flux densities with these measurements. The pulse profiles from Dec.~14 showed a complex structure of a short, strong peak preceded by a less intense and longer in duration precursor, as reported at other frequencies in \cite{Levin2019}. In turn, the precursor peak is not visible on subsequent observations as shown in Fig.~\ref{fig:magnetar}. 


Magnetars can display strong linear polarization \citep{Camilo:2007rm,Torne:2016zid} and their study will benefit from those kind of measurements at IAR. Daily observations allow to follow glitches and the search for a binary magnetar. The increase in sensitivity of the antennas will allow us to investigate possible patterns of stabilization of individual pulses. This search would benefit of an appropriate machine learning clustering algorithm running on our data bank.

\subsection{Glitches and young pulsars}
\label{sec:glitches}

Although pulsars have extremely stable periods over time, some young pulsars are prone to have glitches: sudden changes in their period due to interior changes in the star. Discovered 50~years ago, nowadays almost 200~pulsars are known to glitch \citep{Manchester:2018jhy}. Southern \citep{Yu:2012mp} and northern \citep{Espinoza:2011pq,Fuentes:2017bjx} based surveys provide comprehensive catalogs\footnote{\url{http://www.atnf.csiro.au/people/pulsar/psrcat/glitchTbl.html}\\ \url{http://www.jb.man.ac.uk/pulsar/glitches/gTable.html}}.

Magnetars present the largest relative glitches in frequency ($\nu=P^{-1}$) with $\Delta\nu/\nu \sim10^{-6}$ while, for young pulsars $\Delta\nu/\nu \sim10^{-7} - 10^{-8}$, and for MSPs $\Delta\nu/\nu \sim10^{-11}$. The increase in frequency is generally followed by an exponential decrease that, lasting 10 to 100~days, tries to recover the pre-glitch period, though a permanent change remains.

While the steady slow down of the pulsar spin is most likely produced by magnetic braking taking place  outside the neutron star, glitches are thought to be produced by the sudden coupling of a fast rotating superfluid core with the crust, transferring to it some of the core's angular momentum and hence producing the decrease of the pulsar period. Details of the modeling of such coupling have been challenged \citep{Andersson:2012iu,Chamel:2012ae,Piekarewicz:2014lba} and are a matter of current research \citep[see][for a review on models of pulsar glitches]{Haskell:2015jra}.

\begin{figure}
  \centering
    \includegraphics[width=1.0\linewidth]{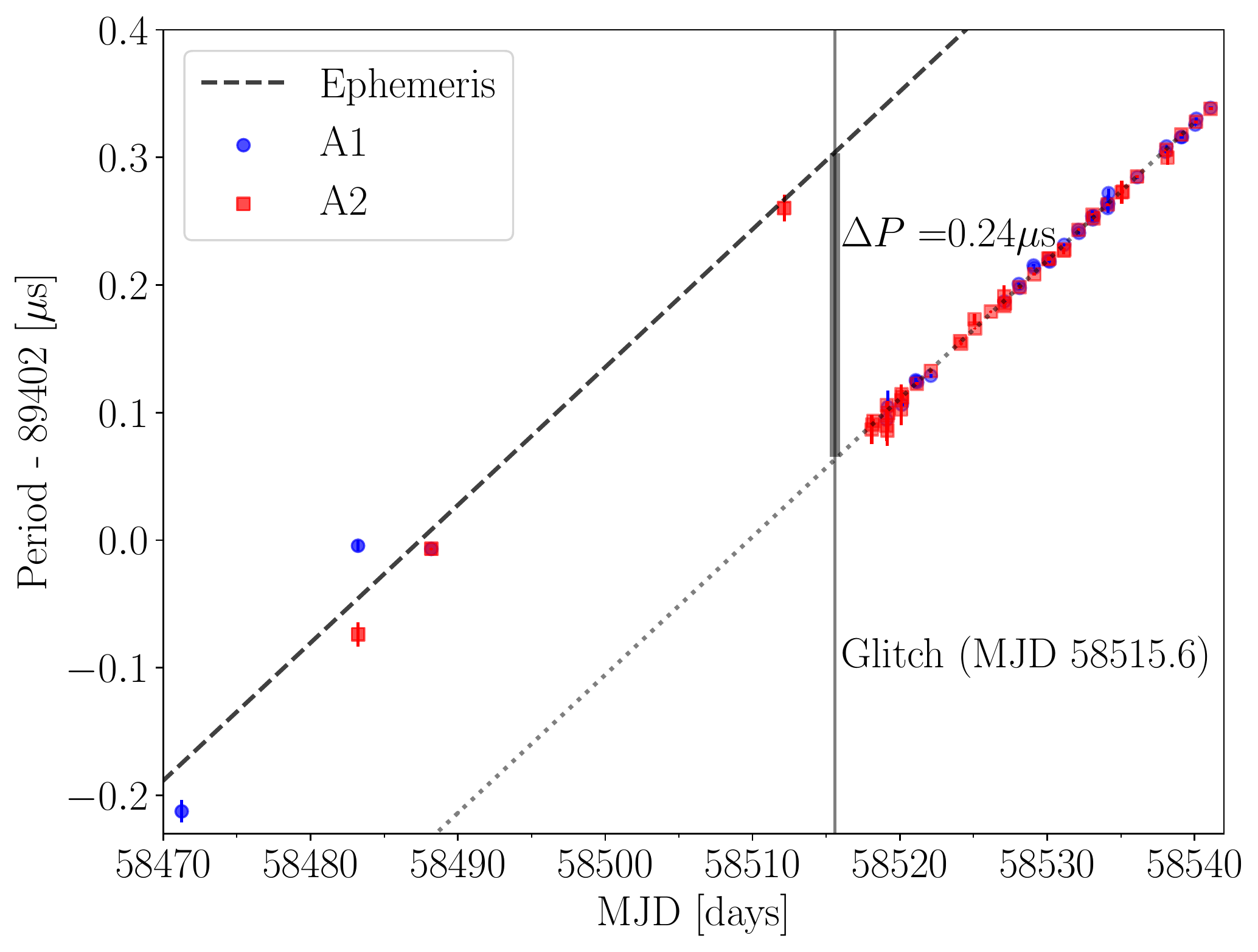}
    \caption{Vela's glitch on February 1st. 2019 as measured at IAR in the 1400~MHz band. Ephemeris for the pre-glitch epoch were taken from \cite{Sarkissian2017} and the glitch date from \cite{atel_fermi_vela}. The errorbars at one $\sigma$ level are taken directly from the output of PRESTO and are likely to represent an underestimation. The dispersion is larger for short observations (10--15~min) than for long ones ($> 60$~min). We highlight that at MJD~58487 the data points from A1 and A2 overlap and are hardly distinguishable.}
\label{fig:VelaGlitch}
\end{figure}

The Vela Pulsar (PSR B0833$-$45/J0835$-$4510) is one of the most active pulsars in terms of glitching, counting 20 in the last 50 years. The latest glitch occurred recently, around MJD~58515, and was reported by \cite{atel_newglitch_vela}. We briefly summarize the radio timing observations performed at IAR of this event, which we first reported in \cite{atel_vela}. As part of the commissioning and developing stage, regular observations of Vela with both antennas were restarted by the end of January 2019 after one month of inactivity. We observed Vela on Jan. 29 (MJD~58512.14) and obtained $P_\mathrm{bary}=89.402260(7)$~ms, consistent with the available ephemeris before the glitch. After the new glitch was reported, we started a daily follow-up of the event starting on Feb. 04 (MJD~58518.15). The monitoring initially consisted of a combination of short (10--15~min) and long (60--220~min) observations. The reconstruction of the post-glitch ephemeris, shown in Fig.~\ref{fig:VelaGlitch}, yields a period jump of $\Delta P \sim -0.241~\mu$s, equivalent to a frequency jump of $3.0\times10^{-5}$~Hz, that is consistent with the value estimated by \cite{atel_fermi_vela}, within 7\% error. 
%

Other glitching pulsars currently within the reach of IAR observatory and their properties are given in Table~\ref{tab:glt} \cite[adapted from ATNF\footnote{\url{http://www.atnf.csiro.au/research/pulsar/psrcat/}} Catalog,][]{Manchester:2004bp}. 
\begin{table}
\centering
\caption{Potentially observable glitching pulsars from IAR. $P_0$ is the barycentric period of the pulsar, $S_{1400}$ the mean flux density at 1400~MHz, $W_{50}$, width of pulse at 50\% of peak (ms), $D$ is the distance based on the \cite{Yao_2017} electron density model, and $N_\mathrm{G}$ is the number of glitches reported for the pulsar. Pulsars marked with a $^\dagger$ are already being monitored at IAR.}\label{tab:glt}
\begin{tabular}{lccccc}
\hline
JName        &   $P_0$   & $S_{1400}$    &   $W_{50}$\tablefootmark{a}   &  $DM$    & $N_\mathrm{G}$ \\
            &   [s]     &  [mJy]        &   [ms]   &    [pc cm$^{-3}$]&   \\
\hline
J0835$-$4510$^\dagger$   &   0.08933   & 1050    & 1.4       &     67.97  &   19  \\
J1644$-$4559$^\dagger$   &   0.45506   &  300.0   & 8.0       & 478.8  &   3   \\    
J1731$-$4744$^\dagger$   &   0.82983   &  27.00    & 17.1         &    123.056  &   7   \\
J0742$-$2822$^\dagger$   &   0.16676   &  26.00    & 4.2         & 73.728  &   7   \\
J1721$-$3532$^\dagger$   &   0.28042   &  16.80    & 29.8      &    496.0  &   1   \\
J1709$-$4429$^\dagger$    &   0.10246   &  12.10    & 5.7    &     75.68  &   1   \\
J1803$-$2137$^\dagger$   &   0.13367   &   9.60    &  13.1 &  233.99  &   5   \\
J1818$-$1422   &   0.29149   &   9.60    & 17.1        &     622.0  &   1   \\
J1705$-$1906   &   0.29899   &  9.30     & 154.2        &     22.907  &   1   \\
J1048$-$5832$^\dagger$   &   0.12367   &  9.10     & 4.8   &   128.679  &   3   \\
J1740$-$3015$^\dagger$   &   0.60689   &  8.90     & 2.4        &    151.96  &   36  \\
J1757$-$2421   &   0.23411   &   7.20    & 9.6     &     179.454  &   1   \\
J1801$-$2304   &   0.41583   &   7.00    & 128.3      &     1073.9  &   9   \\
J1705$-$3423   &   0.25543   &   5.30    & 11.7       &     146.36  &   3   \\
J1539$-$5626   &   0.24339   &  5.00     & 7.6        &     175.85  &   1   \\
J1826$-$1334   &   0.10149   &   4.70    & 5.9        &     231.0  &   6   \\
J1302$-$6350   &   0.04776   &   4.50    &  23.4   &    146.73  &   1   \\
J1739$-$2903   &   0.32288   &   4.50    & 6.0        &     138.55  &   1   \\
J1328$-$4357   &   0.53270   &   4.40    &   10.4      &     42.0  &   1   \\
J1730$-$3350   &   0.13946   &   4.30    &    7.1     &     261.29  &   2   \\
J1614$-$5048   &   0.23169   &   4.10    &    8.4     &     582.4  &   1   \\
J1809$-$1917   &   0.08275   &   2.80    & 17.0      &     197.1  &   1   \\
J1341$-$6220   &   0.19334   &   2.70    & 7.7     &  719.65 &   12  \\
J0758$-$1528   &   0.68227   &  2.60     &  3.4    &  63.327  &   1   \\
J1835$-$1106   &   0.16591   &    2.50   & 3.9 &   132.679  &   1   \\
J1141$-$6545   &   0.39390   &   2.40    & 4.4  &   116.080  &   1   \\
J1743$-$3150   &   2.41458   &   2.10    & 42.3  &  193.05  &   1   \\
\hline
\end{tabular}
\tablefoot{\tablefoottext{a}{Quoted values are indicative only, as the width of pulse at 50\% of peak is a function of both observing frequency and time resolution.}}
\end{table}


IAR's program of pulsar observations considers their follow up for up to 4 hours per day. Hence, there is a chance that during this collected data a glitch could be observed "live". In the case of the very bright Vela pulsar it will be possible to observe single pulses. In order to do so, we need to achieve a higher sensitivity. Ongoing tests suggest that IAR antennas are currently capable of detecting the Vela pulsar with an integration time as small as 0.4~s (i.e., 5 pulses added) with a significance greater than 5$\sigma$. With the future improvements in the antennas receivers (Sect.~\ref{sec:upgrades}), which include a combination of broader bandwidth and reduction of system temperature, it will be possible to study the dynamical spectra of single pulses.

\subsection{Fast-Radio-Burst observations}
Fast radio bursts (FRB) are intense bursts of radio emission that can reach flux densities of 10's of Jy, have a duration of milliseconds and exhibit the same characteristic dispersion sweep in frequency as radio pulsars.
The dispersion measure of these radio bursts together with models of the intergalactic medium suggest that the FRBs may come from as distant as 1~Gpc.
The first FRB 010724 was reported by \cite{Lorimer:2007qn} looking at archival survey data from the Parkes observatory. The single dispersed pulse, followed a $\sim \nu^{-2}$-law, had a width of less than 5~ms and a dispersion measure consistent with a distance of $D\sim 400$~Mpc, which clearly placed the source at extragalactic distances.
Since then over 72 FRBs have been reported \citep{Ravi:2019iop}, but most notably there are cases in which they have been observed to repeat. The first one, FRB 121102, was observed by the Arecibo and then Green bank and other radio telescopes, and also in X-rays \citep{Spitler:2016dmz,Scholz:2016rpt}, in 2015. 
The second repeating FRB 180814.J0422+73, has recently been reported by the CHIME/FRB team \citep{Amiri:2019bjk} (while most of the FRBs have been found in the 1400 MHz band, CHIME found its 13 FRBs in the 400--800~MHz band).
FRB repeaters observations are very important in order to better understand the physics underneath, and possibly to rule out models leading to single catastrophic events. Statistical analysis presented by \citet{Ravi:2019iop} suggests that FRBs are more likely produced by repeating sources; thus, radio monitoring for repeating FRBs seems feasible and it is crucial to probe the nature of these sources.
Notably, during the writing of this manuscript, eight new FRB repeaters have been reported by CHIME/FRB \citep{2019arXiv190803507T}.

Crucial to understand the sources of such FRBs is the identification of an optical or X-ray counterpart. For FRB 121102 a supernova type I in a dwarf galaxy at $z=0.19$ has been found in a coincidence position \citep{Chatterjee:2017dqg,Bassa:2017tke,Eftekhari:2019jah} using VLA observations.
There are other models for FRBs ranging from SGRs, merging of white dwarfs or neutron stars, collapsing supra-massive neutron stars to evaporating primordial black holes and cusps of superconducting cosmic strings.
For a thorough current review of the FRB field see \cite{Cordes:2019cmq}.

A southern hemisphere search for repeating FRBs has been started by the Australian SKA pathfinder \citep{Bhandari:2019med}; a new FRB 180924 event was reported already in \cite{Bannister:2019iju} to occur at $\sim$4~kpc from the center of a luminous Galaxy at $z=0.32$.

There are essentially two strategies to follow to observe FRBs with IAR antennas. One is to observe areas of the sky we expect to produce FRBs, like nearby galaxy clusters.
The second strategy is to follow known FRBs to look for a repeater. With rates up to 10\,000 FRB/sky/day as reported in Table 3 of \cite{Petroff:2019tty} and two observing antennas with large available observing time, 
we can do a first estimation of the rate of potentially detectable FRBs by IAR as $\dot{N}_\mathrm{IAR} \sim \dot{N}_\mathrm{FRB} \times \Delta \Omega/(4\pi)$, with $\Delta \Omega \approx \pi \theta_\mathrm{FWHM}^2/4$. This first guess suggests a likely detection after $\sim 500$~h of accumulated observations. However, considering the lower sensitivity of IAR antennas with respect to other observatories such as Parkes, a more realistic perspective is to have a serendipitous detection after $\sim5\,000$~h of observations, although a large uncertainty could be added due the the unknown luminosity function of the FRBs.

We note that observations of transient events at IAR will have a great synergy with other local observational facilities, such as the 47-cm optical telescope Telescopio Rafael Montemayor (TRM), to be installed in Argentina in early 2020. 
This telescope will be fully automated and it will have a $1\degr \times 1\degr$ field of view, which makes it highly suited to monitor target of opportunity sources, such as counterparts of FRBs. 
Also note that the MASTER-OAFA robotic telescope (OAFA observatory of San Juan National University, Argentina) \footnote{\url{http://master.sai.msu.ru/masternet/}} is already working on fast response FRB follow ups.

The biggest problem faced when looking for fast radio bursts is that we do not know when and where one will flash next. The radio sky has to be observed until one of those events is captured; it is thought that thousands of FRBs occur per day, so serendipitous detections are feasible. This implies that large amounts of raw radio-astronomical observations have to be accumulated and analyzed. Observations of 400~MHz of bandwidth, scanned every 100~$\mu$s and in frequency bins of 0.2~MHz lead to a typical amount of 300~GB per hour of observation, that is potentially 7.2~TB per day and 2.6~PB per year. A minimal requirement of bandwidth of 200~MHz and frequency bins of 0.4~MHz would add up to 0.65~PB per year. Considering that with the IAR observatory we can track objects in the sky for up to 4 hours per day, any single observation can contain FRB data and one can think of applying a machine learning algorithm to look for those kind of patterns in the accumulated raw data. As an example of the potential of such archival searches, recently \cite{Zhang:2019xzf} found a new FRB (FRB 010312) in the original archival data set of the first (FRB 010724) detection, almost two decades later.

\subsection{Interstellar medium scintillation}


Scintillation of radio signals from pulsars is significantly affected by scattering in the turbulent interstellar medium. Understanding and accurately modeling this source of noise is crucial to improve the sensitivity of the pulsar timing arrays to detect gravitational waves from SMBHBs.

A classic study based on a daily analysis of the flux density at 610~MHz of 21 pulsars for 5 years by \cite{Stinebringetal} yielded valuable information of the flux variations caused by interstellar effects through the mechanism of "refractive scintillation".


Similar detailed studies \citep{Reardon:2019mjd,ReardonPhDThesis} of the long-term changes in the diffractive scintillation pattern of the binary pulsar PSR J1141$-$6545 during 6 years allowed to improve the determination of the binary parameters and to give an estimate of its proper motion. 


We can perform similar studies from IAR in the southern hemisphere, either by systematic studies of a selected set of pulsars via daily monitoring or by use of the information already collected for targeted pulsars for other projects, like for J0437$-$4715. Our observations would supplement those in other southern observatories, in particular those from Australia, providing a double cadence, every nearly 12 hours.


\subsection{Tests of Gravity with Pulsar timing}

Pulsars in a binary system can be used as tests of alternative theories of gravity. They have been extensively used in the post-Newtonian parametrization approach to weak gravitational fields \citep[see][for a review]{Will:2018bme}. In particular one can test General Relativity versus modified theories
\citep[see][for a current review]{Will:2018bme,Renevey:2019jrm}.

Notably, in a recent paper \cite{Yang:2016rig} proposed to use scintillation measurements of PSR B0834+06 to test predictions of alternative theories of gravity. This work could be made extensive to J0437$-$4715 and other accurately measured MSPs.

Pulsar timing arrays can constrain alternative theories of gravity via its strong sensitivity to gravitational waves polarizations, particularly the longitudinal polarization of vector and scalar modes \citep{2008ApJ...685.1304L,daSilvaAlves:2011fp,Chamberlin:2011ev,Cornish:2017oic}. SKA, with its extraordinary sensitivity will be able to be at the forefront of those tests of gravity \citep[e.g.][]{Bull:2018lat}.

Pulsar timing have also been proposed as a mean to detect and to shed some light on the constituents of dark matter \citep[e.g.][]{Khmelnitsky:2013lxt}.




\section{Conclusions}\label{sec:conclusions}

We have shown that with IAR's upgrades we can now perform 4 hours of continuous observations of bright pulsars of the southern hemisphere with a daily cadence for very long periods of time. This gives an estimate of 1000 hours/year/pulsar/antenna.  These capabilities allow us to contribute to the international pulsar timing array efforts, and to follow up targeted MSPs of the LIGO-Virgo collaboration. We have also been able to already monitor magnetar activity and pulsar glitches. Notably IAR's location provides a 12 hour complementary window with respect to australian Parkes' and over 5 hours of MeerKAT observations, and thus can uniquely cover transient phenomena such as FRBs or live-glitches.

In order to increase the number of pulsars and the accuracy of measurements for those projects we are developing a tighter calibration of the intensity and polarization observations (particularly important for magnetars measurements), automation of observations (for applications to interstellar scintillation measurements and glitches surveys, for instance), and increase of the RF bandwidth by an order of magnitude (with new/additional USRP plates matching IAR site RFI frequency windows) and data downloading bandwidth (RF through fiber optics) for applications to FRB searches, for instance.

We plan to store all our raw data and make it available to the astronomical community for archival post-processing and further analysis, both for continuous gravitational waves searches and interstellar scintillation studies as well as for transient phenomena such as FRBs, (mini-)glitches, magnetars, and other unexpected astrophysical phenomena.

Future perspectives, in addition to the upgrades to IAR's radio antennas, include observation time in the two 35~m antennas located in Mendoza 
\footnote{\url{https://www.esa.int/Our_Activities/Operations/Estrack/Malarguee_-_DSA_3}} and in Neuquen
\footnote{\url{https://www.argentina.gob.ar/ciencia/conae/centros-y-estaciones/estacion-cltc-conae-neuquen}}, Argentina. Regarding optical counterparts, we plan to perform simultaneous observations with the facilities already present at San Juan \footnote{\url{https://casleo.conicet.gov.ar/hsh/}}, Argentina.

The detection of gravitational waves from supermassive black hole mergers can yield fundamental insight into their astrophysical formation and growth scenarios as well as the structure and evolution of the universe, and, ultimately, provide crucial tests for classical theories of gravity.
IAR is now ready to make its modest contributions towards this goal.


\begin{acknowledgements}
The authors thank various members of the IAR's technical staff for their work, as well as numerous members of the NANOGrav and LIGO-Virgo collaborations for very valuable discussions, in particular D.Keitel and J.Romano for reading our original manuscript. Noteworthy have also been the additions and corrections suggested by an anonymous referee.
Part of this work was supported by the National Science Foundation (NSF) from Grants No. PHY-1912632, No.\ PHY-1607520, and No.\ PHY-1726215. 
FG and JAC acknowledge support by PIP 0102 (CONICET). 
This work received financial support from PICT-2017-2865 (ANPCyT). 
JAC was also supported by the Agencia Estatal de Investigaci\'on grant AYA2016-76012-C3-3-P from the Spanish Ministerio de Econom\'ia y Competitividad (MINECO) and by the Consejer\'ia de Econom\'ia, Innovaci\'on, Ciencia y Empleo of Junta de Andaluc\'ia under research group FQM-322, as well as FEDER funds. 
\end{acknowledgements}


\bibliographystyle{aa}
\bibliography{biblio}

\end{document}